\documentclass[a4paper,11pt]{article}
\pdfoutput=1 

\usepackage{jheppub} 

\usepackage[T1]{fontenc} 

\usepackage{soul}
\usepackage[inline]{enumitem}
\usepackage[utf8]{inputenc}
\usepackage[normalem]{ulem}
\usepackage{graphicx}
\usepackage{dcolumn}
\usepackage{bm}
\usepackage{color}
\usepackage[dvipsnames]{xcolor}
    
\usepackage{url}
\usepackage{xspace}
\usepackage{slashed}
\usepackage{multirow,bigstrut}
\usepackage{mathrsfs} 
\usepackage{relsize,amsmath}
\usepackage[geometry]{ifsym}
\usepackage{amssymb}
\usepackage{pifont}
\usepackage{physics}
\usepackage{ulem}

\usepackage{cleveref}

\usepackage{rotating}
\usepackage{ragged2e}

\usepackage{fontawesome} 
\definecolor{blue-violet}{rgb}{0.33, 0.17, 0.89}

\newcounter{CommentCount}
\definecolor{MH}{rgb}{0.0,0.6,9}
\definecolor{palatinate}{rgb}{0.494, 0.192, 0.482}

\definecolor{teal}{HTML}{008080}

\usepackage{siunitx}

\DeclareSIUnit \s {\second}
\DeclareSIUnit \ns {\nano\second}
\DeclareSIUnit \mus {\micro\second}
\DeclareSIUnit \ms {\milli\second}
\DeclareSIUnit \MB {\mega\byte}
\DeclareSIUnit \GB {\giga\byte}
\DeclareSIUnit \TB {\tera\byte}
\DeclareSIUnit \PB {\peta\byte}
\DeclareSIUnit \Mbps {\mega\bit/\s}
\DeclareSIUnit \Gbps {\giga\bit/\s}
\DeclareSIUnit \Tbps {\tera\bit/\s}
\DeclareSIUnit \Pbps {\peta\bit/\s}
\DeclareSIUnit \kton {\kilo\tonne} 
\DeclareSIUnit \kt {\kilo\tonne}
\DeclareSIUnit \Mt {\mega\tonne}
\DeclareSIUnit \eV {\electronvolt}
\DeclareSIUnit \keV {\kilo\electronvolt}
\DeclareSIUnit \MeV {\mega\electronvolt}
\DeclareSIUnit \GeV {\giga\electronvolt}
\DeclareSIUnit \TeV {\tera\electronvolt}
\DeclareSIUnit \PeV {\peta\electronvolt}
\DeclareSIUnit \EeV {\exa\electronvolt}
\DeclareSIUnit \m {\meter}
\DeclareSIUnit \cm {\centi\meter}
\DeclareSIUnit \in {\inchcommand}
\DeclareSIUnit \km {\kilo\meter}
\DeclareSIUnit \kV {\kilo\volt}
\DeclareSIUnit \kW {\kilo\watt}
\DeclareSIUnit \MW {\mega\watt}
\DeclareSIUnit \MHz {\mega\hertz}
\DeclareSIUnit \mrad {\milli\radian}
\DeclareSIUnit \year {years}
\DeclareSIUnit \POT {POT}
\DeclareSIUnit \sig {$\sigma$}
\DeclareSIUnit\parsec{pc}
\DeclareSIUnit\lightyear{ly}
\DeclareSIUnit\foot{ft}
\DeclareSIUnit\ft{ft}
\DeclareSIUnit \ppb{ppb}
\DeclareSIUnit \ppt{ppt}
\DeclareSIUnit \samples{S}
\DeclareSIUnit \pe{PE}
\DeclareSIUnit \T{T}

\newcommand{\enu}{\E_\enu}

\RequirePackage[normalem]{ulem}

\title{\boldmath Neutron stars can shine a light on elusive lepton-flavor-violating dark matter}

\author[a]{Hooman Davoudiasl,}
\author[b]{Jaime Hoefken Zink,}
\author[b,c]{Sebastian Trojanowski}

\affiliation[a]{High Energy Theory Group, Physics Department, Brookhaven National Laboratory, Upton, NY 11973, USA}
\affiliation[b]{National Centre for Nuclear Research, Pasteura 7, Warsaw, PL-02-093, Poland}
\affiliation[c]{Astrocent, Nicolaus Copernicus Astronomical Center Polish Academy of Sciences, ul.~Rektorska 4, 00-614, Warsaw, Poland}

\emailAdd{hooman@bnl.gov}
\emailAdd{jaime.hoefkenzink@ncbj.gov.pl}
\emailAdd{sebastian.trojanowski@ncbj.gov.pl}

\abstract{We investigate a scenario in which dark matter (DM) poses a challenge to conventional direct and indirect detection, making it much more  elusive than  typical candidates.  We focus on thermally produced DM that couples to electrons and muons via a lepton-flavor-violating (LFV) axion-like particle (ALP). Given the DM kinematics and lack of muon targets on Earth, direct detection would be infeasible. Indirect detection of our DM candidate is also hampered by the dominance of $p$-wave annihilation. However, we demonstrate that neutron stars (NS) can serve as probes of such a scenario, through future dedicated observational campaigns. Infalling DM is accelerated to semi-relativistic velocities, triggering inelastic $\chi e \leftrightarrow \chi \mu$ scattering off both electrons and muonic targets within the NS. We show that ``flavor blocking'' -- the kinematic suppression of LFV interactions at low energies -- prevents DM thermalization with the cold neutron star, enabling efficient $p$-wave annihilations. The resulting NS surface temperatures ($T_s \gtrsim 2 \times 10^3~\mathrm{K}$) offer a possible signature for future infrared searches, probing thermal relics beyond the reach of direct, indirect, and accelerator experiments.}

\begin{document}
\maketitle
\flushbottom

\section{Introduction}

The microscopic nature and origin of dark matter (DM) remain unknown. A leading hypothesis is that DM was produced in thermal equilibrium in the early, radiation-dominated universe. This paradigm of thermal DM production has motivated numerous experimental efforts, with current and next-generation DM detectors aiming to explore this possibility fully~\cite{Roszkowski:2017nbc,Billard:2021uyg}. However, as experimental limits tighten, it is clear that many viable thermal DM scenarios reside in regimes that are notoriously difficult to probe using standard methodologies.

This lack of sensitivity arises from the fundamental differences between the conditions of the early universe and those of the present day. While thermal freeze-out occurred at semi-relativistic velocities, contemporary direct detection (DD) and indirect detection (ID) experiments are primarily sensitive to the non-relativistic limit. Furthermore, the thermal relic abundance may be determined by DM interactions with Standard Model (SM) species that are unstable or kinematically inaccessible in terrestrial laboratories. To fully explore the thermal DM parameter space, it is essential to extend the scope of traditional DM searches and to look beyond them~\cite{Boveia:2022syt}. Astrophysical signatures play a key role in this endeavor, particularly for beyond the Standard Model (BSM) scenarios that are otherwise difficult to probe~\cite{AlvesBatista:2021eeu}.

Neutron stars (NSs) have been recognized as a particularly compelling laboratory to test DM interactions through expected DM-induced NS heating~\cite{Kouvaris:2007ay,Bertone:2007ae,Kouvaris:2010vv,deLavallaz:2010wp,Bertoni:2013bsa,Baryakhtar:2017dbj,Bramante:2017xlb,Raj:2017wrv,Bell:2018pkk,Acevedo:2019agu,Bell:2020jou,Bell:2020obw,Garani:2020wge,Dasgupta:2020dik,Anzuini:2021lnv,DeRocco:2022rze,Coffey:2022eav,Nguyen:2022zwb,Bell:2023ysh,Bramante:2023djs,Zhang:2023vva,Su:2024flx,Ema:2024wqr} or excessive DM accretion leading to NS collapse and black hole formation~\cite{Goldman:1989nd,McDermott:2011jp,Bramante:2014zca,Garani:2018kkd}. Their extreme gravitational fields accelerate infalling DM to semi-relativistic speeds, effectively recreating the kinetic conditions of the early universe. Moreover, their dense interiors, containing high number densities of both electrons and muons~\cite{Haensel:2007yy}, allow for the exploration of dark sectors that couple preferentially to leptons~\cite{Bell:2019pyc,Garani:2019fpa,Joglekar:2019vzy,Joglekar:2020liw,Bell:2020lmm}. Future observations of old, cold NSs, where DM-induced heating could be the dominant thermal source, can provide a sensitive probe for interactions that are otherwise beyond the reach of traditional detectors.

In this study, we investigate a particularly challenging case: lepton flavor violating (LFV) dark matter. We focus on a scenario where DM couples to the SM via an $e$-$\mu$ coupling that allows $\chi e\leftrightarrow \chi \mu$ transitions. This framework is characterized by several features that render it nearly invisible to standard searches:
\begin{enumerate}
\item Direct detection is suppressed by the absence of tree-level couplings to quarks.
\item Kinematic blocking prevents scattering on leptonic targets; because DM in the local halo is non-relativistic, it lacks the kinetic energy required to overcome the mass threshold for $e\to\mu$ transitions in terrestrial detectors.
\item Indirect detection is hampered if the interaction is mediated by a pseudoscalar, such as axion-like particle (ALP), which typically leads to $p$-wave suppressed annihilation cross-sections at the low velocities found in galactic halos.
\end{enumerate}

We demonstrate that the unique environment of a neutron star effectively overcomes these barriers. The gravitational boost provides sufficient energy for the $\chi e\to \chi\mu$ upscattering, while the intrinsic muon population within the NS interior enables the inverse $\chi\mu\to\chi e$ transition. Notably, we find that the kinematics of LFV scattering result in a \textsl{flavor-blocking mechanism}, which prevents the DM population from reaching full thermal equilibrium with the stellar temperature. This leads to a ``warmer'' DM distribution within the star that facilitates efficient annihilation heating, even in the case of $p$-wave suppressed models.

Using an ALP -- which could naturally be a light particle -- as a concrete yet flexible framework, we show that future observations of old NSs with temperatures $T\sim 2000~\textrm{K}$ can provide leading constraints on LFV thermal targets. This study represents the first analysis of LFV DM in the context of neutron stars, offering a path to probe dark sectors that remain largely inaccessible to terrestrial and traditional astrophysical searches, with the possible exception of future high-energy lepton colliders.

\section{The LFV model}

We illustrate the LFV dark sector physics using a portal involving light axion-like particles~\cite{DiLuzio:2020wdo,Choi:2020rgn,Bauer:2020jbp}. ALPs can naturally emerge as light pseudo-Nambu-Goldstone bosons from global symmetries spontaneously broken at high energy scales. They can mediate interactions between DM and the SM. In this framework, ALP couplings need not respect the SM flavor structure and could even play a role in generating it, e.g., through familons~\cite{Davidson:1981zd,Wilczek:1982rv} or flavons~\cite{Bauer:2016rxs}. Furthermore, LFV couplings of ALPs can be radiatively generated~\cite{Choi:2017gpf}, providing a natural theoretical basis for LFV interactions of DM. By employing the ALP as a mediator, we establish a concrete yet flexible framework that captures the essential physics of NS heating from LFV DM interactions without being obscured by excessive BSM complexity.

We assume that the dark sector manifests at lower energies primarily through these ALP-induced couplings. Specifically, we consider a model where the ALP, denoted as $a$, couples predominantly through LFV interactions with muons and electrons,
\begin{equation}
\label{eq:lag_alp_e_mu}
\mathcal{L} \supset - i g_{e \mu} a \overline{e} \left[ \sin{\phi} + \cos{\phi} \gamma^5 \right] \mu + \mathrm{h.c.}
\end{equation}
In the following, we will consider a purely pseudoscalar coupling by setting $\phi = 0$. We note that the precise value of this angle plays a minor role in the semi-relativistic regime of scattering within NSs.

Such LFV couplings can be inherited from the interactions of new heavy fields at a scale $\Lambda$, such that $g_{\mu e} = C_{\mu e}\,m_\mu/\Lambda$ and $C_{\mu e}\sim 1$~\cite{Batell:2024cdl}. Requiring that these new fields be accessible at colliders, {\it i.e.}, $\Lambda\lesssim 10~\textrm{TeV}$, provides a target of $g_{\mu e}\gtrsim 10^{-5}$, which can only be partially probed in future ALP searches~\cite{Dev:2017ftk,Endo:2020mev}. Testing even lower couplings requires more indirect probes. In particular, the couplings in \cref{eq:lag_alp_e_mu} can induce rare muon decay rates far exceeding SM predictions. However, for ALP masses $m_a>m_\mu$, where $m_\mu$ is the mass of the muon, these bounds often require both LFV and LFC couplings~\cite{Bauer:2019gfk,Cornella:2019uxs,Calibbi:2020jvd}. Exploring ALP scenarios where the LFC couplings are strongly suppressed is considerably more difficult and may necessitate novel experimental techniques~\cite{Gninenko:2001id,Gninenko:2018num,Davoudiasl:2021haa,Cheung:2021mol,Bauer:2021mvw,Davoudiasl:2021mjy,Haghighat:2021djz,Ema:2022afm,Radics:2023tkn,Davoudiasl:2024vje,Calibbi:2024rcm,Batell:2024cdl} or astrophysical probes~\cite{Calibbi:2020jvd,Zhang:2023vva,Li:2025beu,Huang:2025rmy,Huang:2025xvo}. While non-vanishing LFC couplings may still naturally be induced in UV-complete scenarios and would lead to more standard phenomenology, we treat this pure LFV regime as an extreme benchmark to probe DM detection limits when such standard interactions are suppressed.

We further introduce ALP couplings to DM, which we assume consists of a Dirac fermion $\chi$,
\begin{equation}
\label{eq:lag_alp_dm}
\mathcal{L} \supset - i g_{\chi} a \overline{\chi} \gamma^5 \chi\,.
\end{equation}
Here, the dark coupling constant $g_\chi$ can be, \textsl{a priori}, unrelated to $g_{\mu e}$. However, the couplings can also exhibit a natural scaling, $g_\chi/g_{\mu e} \sim m_\chi / m_\mu$, provided that they originate from a common dark sector at the UV completion scale $\Lambda$.

If the dark coupling constant is not suppressed, $g_\chi \gg g_{\mu e}$, the DM relic density is determined by the dominant secluded annihilation mode, $\bar{\chi}\chi\to aa$. The corresponding cross section is given by~\cite{Armando:2023zwz}
\begin{align}
\langle \sigma  v \rangle_{aa}   \simeq  \frac{6}{x_{\rm f.o.}}  \frac{ g_\chi^4}{ 24 \pi } \frac{ m_\chi^2 (m_\chi^2 - m_a^2)^2}{ (2 m_\chi^2-m_a^2)^4} \left( 1- \frac{m_a^2}{m_\chi^2}  \right)^{1/2}\,,
\label{eq:annsecluded}
\end{align}
where $x_{\rm f.o.}\equiv m_\chi/T_{\rm f.o.}$, at freeze-out temperature $T_{\rm f.o.}$. Here, requiring that the thermal DM abundance equals $\Omega_\chi^{\textrm{th}} h^2 \simeq 0.12$~\cite{Planck:2018vyg} determines $g_{\chi,\textrm{th}}$ as a function of the dark sector masses. For our benchmark scenario, we find that $g_{\chi,\textrm{th}} \simeq 0.1\,\sqrt{m_a/\textrm{GeV}} \gg g_{\mu e}$ within the allowed parameter space. Importantly, this secluded annihilation mode is $p$-wave suppressed, rendering indirect detection signals in the present-day galactic halo strongly suppressed.

While additional $s$-wave annihilations into SM leptons, $\bar{\chi}\chi\to\mu e$, are also possible, this process remains subdominant in the early universe for the small $g_{\mu e}$ values we consider. The relevant cross section is given by~\cite{Batell:2024cdl}
\begin{equation}
\langle \sigma  v \rangle_{\mu e}   \simeq \frac{g_{\mu e}^2  g_\chi^2}{ 16 \pi } \frac{(4 m_\chi^2 - m_\mu^2)^2}{m_\chi^2 (4 m_\chi^2-m_a^2)^2}\ .
\label{eq:annvisible}
\end{equation}
At present times, these $s$-wave annihilations introduce additional DM indirect detection (ID) bounds. However, as shown below, future observations of old NSs could yield significantly stronger constraints on $g_{\mu e}$. Collectively, the above features reinforce the elusive nature of the LFV thermal relic with the ALP mediator, as the combination of kinematic thresholds and velocity-suppressed interactions often places the dark sector effectively beyond the reach of traditional searches, positioning neutron stars as a promising probe for constraining this scenario.

Finally, we emphasize that the utility of neutron stars as a probe for LFV is specifically centered on $e$-$\mu$ transitions. By contrast, LFV couplings involving the $\tau$ lepton ($e$-$\tau$ or $\mu$-$\tau$) remain inaccessible in the context of NS heating. This arises from two factors: first, unlike electrons and muons, $\tau$ leptons are not present as targets within the NS interior; second, even with the significant gravitational boost provided by the star, infalling DM particles do not typically attain sufficient kinetic energy to overcome the high mass threshold required for upscattering, $m_\tau\simeq 1.777~\textrm{GeV}$~\cite{ParticleDataGroup:2024cfk}. Consequently, while NS heating is highly sensitive to $e$-$\mu$ dark sector couplings, it will act as a complementary probe to LFV tests in the $\tau$ sector.

\section{LFV dark matter capture} In the vicinity of the NS, DM particles are gravitationally attracted and boosted, leading to DM-induced heating due to their scatterings in the star and possible subsequent annihilations. The DM capture rate as seen by an observer far from the star is given by~\cite{Bell:2020jou}
\begin{align}
C = \frac{4 \pi}{v_\mathrm{NS}} \frac{\rho_\chi}{m_\chi} \mathrm{Erf} \left(\sqrt{\frac{3}{2}} \frac{v_\mathrm{NS}}{v_d} \right)  \times \int_0^{R_{\rm NS}} \sqrt{g_{rr}} r^2 \frac{\sqrt{1 - g_{tt} (r)}}{g_{tt} (r)} \Omega^- (r) \ \eta(r) dr\,,
\end{align}
where $v_\mathrm{NS} = 230$ km/s is the NS velocity with respect to the local DM halo with the DM density of $\rho_\chi = 0.4~\textrm{GeV}/\textrm{cm}^3$, $v_d = 270$ km/s is the DM velocity dispersion, assuming a Maxwell-Boltzmann distribution, and $g_{tt}(r)$ is the time component of the NS metric dependent on the position in the star, and $\sqrt{g_{rr}}$ is the square root of the radial component of the metric, which comes from the Jacobian of the volume integration. The optical factor $\eta(r)$ accounts for the DM absorption within the NS. The interaction rate $\Omega^-(r)$ is the probability of DM scattering that reduces an outgoing $\chi$'s speed below the local escape velocity at position $r$ within the star, $v_e^2 (r) = 1 - g_{tt} (r)$. These captured DM particles deposit their kinetic energy through subsequent scatterings, heating the NS. For relevant DM masses, a single scattering is typically sufficient for capture. 

We emphasize that the interaction rate $\Omega^-$ is calculated using the proper target densities $n_{e,\mu}(r)$ as determined locally by the equation of state (EoS). Since these densities are defined per unit of proper volume, the integration over the star’s volume must be performed using the proper volume element $dV = \sqrt{g_{rr}}\,r^2\,d\Omega dr$. This ensures that the spatial curvature is correctly accounted for.

For LFV DM interactions, kinetic heating involves two interaction modes, such that $\Omega^- \equiv \Omega^-_{\mu \to e} + \Omega^-_{e \to \mu}$. This dual scattering channel is specific to the NS environment; while terrestrial detectors lack both the energy for electron upscattering and the muon targets for downscattering, the NS provides both. DM particles entering the NS can first interact with electrons. The $\chi e\to\chi\mu$ upscattering becomes allowed at a radial distance $R_{\textrm{eff}} < R_{\textrm{NS}}$, where $R_{\textrm{NS}}$ is the radius of the star. $R_{\textrm{eff}}$ marks the position where the lepton chemical potential and density become sufficiently high to allow DM scattering off relativistic electron targets. At a slightly smaller radius, $R_\mu < R_{\textrm{eff}}$, muons are stable inside the NS, and the $\chi\mu\to\chi e$ scattering mode also becomes possible. Detailed expressions for the interaction rate and optical factor are provided in \Cref{sec:capture_rate}.

\begin{figure}[t]
    \centering
    \includegraphics[width=0.65\textwidth]{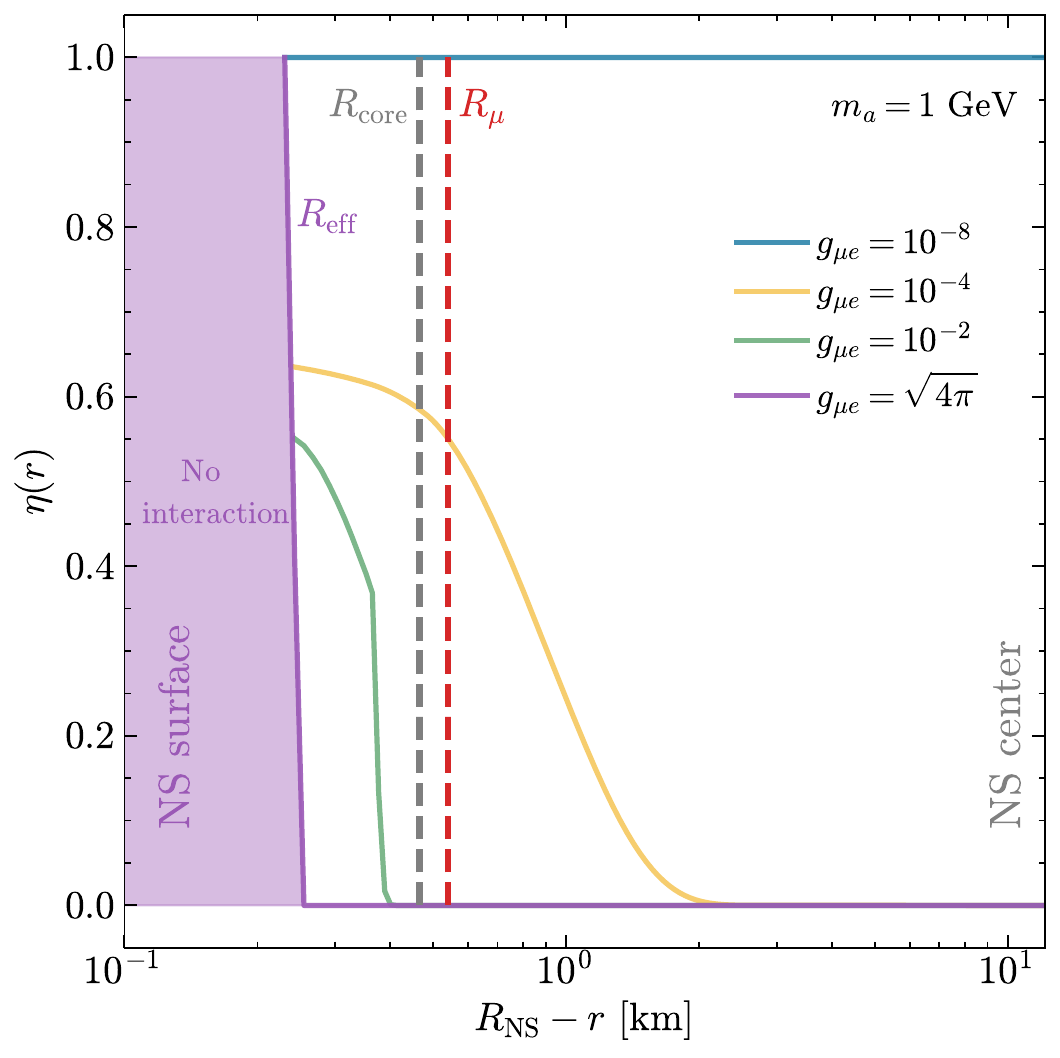}
    \caption{Optical factor, $\eta(r)$, for a $2\,M_\odot$ NS modeled with the BSk25 equation of state. We assume $m_\chi = 2\,m_a$, $m_a = 1~\textrm{GeV}$, and the thermal value of $g_{\chi,\textrm{th}}$. Results are presented for selected $g_{\mu e}$ coupling strengths, as indicated. The purple-shaded region, corresponds to low values of $R_{\textrm{NS}}-r$, where $R_{\textrm{NS}}$ is the NS radius and $r$ is the distance measured from the center of the star. This region marks the regime close to the star's surface where the electron chemical potential and density are too low and their upscattering to muons in the outer NS crust can be neglected. $R_\mu$ denotes the radius of the stable muon region within the star. We indicate the transition between the crust and core of the star by a vertical gray dashed line. The center of the star is toward the right-hand side, where $R_{\textrm{NS}}-r \to 0$. 
\label{fig:opticalfactor}}
\end{figure}

In \cref{fig:opticalfactor}, we present sample optical factors $\eta(r)$ as a function of radial position within an NS, for $m_\chi = 2~\textrm{GeV}$ ($m_a = 1~\textrm{GeV}$) and selected values of the $g_{\mu e}$ coupling. We assume an NS mass of $M_{\textrm{NS}} = 2\,M_\odot$ and the BSk25 equation of state~\cite{Xu:2012uw, Goriely:2013nxa, Perot:2019gwl}, which yields a radius of $R_{\textrm{NS}} = 12.13\,\mathrm{km}$ and central density of $\rho_c = 1.185 \times 10^{15} \, \mathrm{g}/\mathrm{cm}^3$. We also mark the position of the effective radius $R_{\textrm{eff}}$, relevant for this DM mass, and the muon radius $R_\mu$ in the core of the star.

As the figure illustrates, for large couplings, $\eta(r)$ approaches a naive geometric limit, indicating that most DM particles scatter close to the surface of the star at $r\approx R_{\textrm{eff}}$. This contrasts with the optically-thin regime, where $\eta(r)\sim 1$ for diminishing $g_{\mu e}$. In intermediate scenarios, DM particles penetrate deeper into the star, experiencing larger boost factors before interacting in the NS core.

We illustrate this with the yellow line in the plot, obtained for the limiting value of $g_{\mu e} = 10^{-4}$. The initial drop in $\eta(r)$ at $r\approx R_{\textrm{eff}}$ occurs due to partial attenuation of the DM flux as it passes through the NS interior before reaching position $r$. As we discuss below, larger values of $g_{\mu e}$ tend to be excluded by past searches. Hence, we expect DM particles in the considered BSM scenario to typically interact at $r<R_\mu$ and participate in both DM-$e$ and DM-$\mu$ scatterings. 

\section{NS heating} To quantify DM-induced heating, we evaluate the rate of kinetic energy $E_k$ deposition in the NS considering both processes, $\chi (k_1) \, \ell_\alpha (p_1) \to \chi (k_2) \, \ell_\beta (p_2)$, where the 4-momenta are in parentheses and the Greek subscripts refer to the leptons involved, $e$ and $\mu$, 
\begin{align}
\label{eq:E_k_dot}
\dot{E}_k = \frac{4 \pi}{v_\mathrm{NS}} \frac{\rho_\chi}{m_\chi} \mathrm{Erf} \left(\sqrt{\frac{3}{2}} \frac{v_\mathrm{NS}}{v_d} \right) \times \int_0^{R_{\rm NS}} \sqrt{g_{rr}} r^2 \frac{\sqrt{1 - g_{tt} (r)}}{g_{tt} (r)} E_k^- (r) \eta(r) dr\,.
\end{align}
Here, $E_k^- \equiv (E_k^-)_{\mu \to e} + (E_k^-)_{e \to \mu}$, and 
\begin{align}
\label{eq:Ekmin}
(E_k^-)_{\alpha\to\beta} =  &\frac{\zeta}{32 \pi^3 m_\chi} \sqrt{\frac{g_{tt}}{1 - g_{tt}}} \\ &\times \int_{m_\alpha}^{\mu_l} d E_{p_1} E_{p_1} \int_{s_\mathrm{min}}^{s_\mathrm{max}} \frac{s ds}{g_s (m_\alpha) \beta_s (m_\alpha)} \int_{t_\mathrm{min}}^{t_\mathrm{max}} dt \langle\left|\mathcal{M} \right|^2\rangle \ \gamma_s \ E_k \  \Theta (E_{p_2} - \mu_l) \,,\nonumber
\end{align}
where $g_s (m)$ and $\beta_s (m)$ are functions defined in \Cref{sec:capture_rate}, and $\gamma_s \equiv 1 / \sqrt{g_{tt} (R_{\rm NS})}$ is the boost factor on the surface, {\it i.e.}, the boost factor any particle would acquire when coming from rest far from the star to its surface. We note that we measure the energy deposition locally, as seen from the surface of the star, and the $\gamma_s$ factor accounts for the relevant time dilation. This is included in \Cref{eq:E_k_dot} to obtain the energy deposition from the expression for capture rate, since the latter ($C$) represents the amount of particles captured per second as measured far from the star. This rate is modified by gravity on the surface. 

For cold neutron stars, DM capture can be regarded as the only source of heating. The temperature on the star's surface then reads as follows~\cite{Baryakhtar:2017dbj}
\begin{equation}
T_s = \left( \frac{\dot{E}_k}{4 \pi \sigma_\mathrm{SB} R_{\rm NS}^2} \right)^{1/4}\,,
\end{equation}
where $\sigma_\mathrm{SB}$ is the Stefan-Boltzmann constant. Note that $T_s$ refers to the local temperature at the star's surface. For a distant observer, the apparent temperature would be redshifted as $T_\infty = T_s \sqrt{g_{tt}(R_{\textrm{NS}}})$, which for our $2M_\odot$ benchmark corresponds to $T_\infty = 0.72~T_s$. We note that the coldest currently observed neutron stars have temperatures of $\sim 40,000~\mathrm{K}$~\cite{Guillot:2019ugf}. Reaching the temperatures of interest for our scenario, $T_\infty \lesssim 2000~\mathrm{K}$, implies probing thermal luminosities that are approximately five orders of magnitude smaller. Consequently, detecting such faint thermal emission from old NSs is quite challenging for current instruments like the James Webb Space Telescope (JWST)~\cite{Chatterjee:2022dhp} given typical distances to known neutron stars~\cite{Manchester:2004bp}.

Given the low thermal luminosities involved, reaching these targets will likely require dedicated, long-exposure observational campaigns.  Nevertheless, this method may define an objective for future infrared observatories with larger apertures, such as the Extremely Large Telescope (ELT) and its successor missions, cf. the discussion in Refs~\cite{Baryakhtar:2017dbj, Bramante:2023djs}. Crucially, as illustrated in this study, NSs can probe kinematic regimes and scattering targets that are not easily accessible in terrestrial laboratories. Furthermore, identifying DM-induced heating in isolated NSs would provide independent astrophysical evidence of DM interactions beyond the Solar neighborhood. For these next-generation campaigns, old NSs may act as the primary ``detectors'' for elusive dark sectors that are exceptionally challenging to identify via conventional approaches.

DM particles interact with leptons in the star, depositing kinetic energy into it. If a DM particle loses all of its kinetic energy during an interaction at radius $r$ inside the star, the equivalent kinetic energy deposition measured on the NS surface is given by $E^{\mathrm{scat}}_k = \xi_r \left(\gamma_r - 1 \right)  m_\chi$. This relation holds under the assumption that the energy deposited at radius $r$ is efficiently transported to the surface and emitted as radiation. The factor $\xi_r \equiv \gamma_s/\gamma_r$ accounts for the gravitational redshift of this energy between the deposition site and the surface. Here, $\gamma_r$ is the boost factor at $r$. If DM particles deposit most of their kinetic energy close to the NS surface, we find $E^{\mathrm{scat}}_k = \left(\gamma_s - 1 \right)  m_\chi$~\cite{Baryakhtar:2017dbj}. As the DM particle continues to sink towards the center of the star, it keeps depositing the excess kinetic energy it gains in this process~\cite{Bell:2023ysh}. In \Cref{sec:kin_dep}, we show that this additional, continued deposition can be approximately equivalent to considering the initial DM particle interaction to occur at the center of the star, such that $E^{\mathrm{scat}}_k = \xi_0 \left(\gamma_0 - 1 \right)  m_\chi$.

Subsequent DM annihilations close to the NS center deposit additional energy in the star, given by $E_k^\mathrm{ann} = f_\mathrm{ann} \  \xi_0 \  m_\chi$. Here, $f_\mathrm{ann} \equiv 2 \Gamma_\mathrm{ann} / C$ is the fraction of DM particles that annihilate relative to the capture rate. The annihilation rate, $\Gamma_\mathrm{ann} = (1/2)A_\chi\,N_\chi^2$ , depends on the number of DM particles in the star, $N_\chi$. In the parameter space explored in this work, $f_\mathrm{ann} = 1$ for an NS age of $t \geq 10^4$ yr. Therefore, we find the total kinetic energy deposition
\begin{equation}
E_k = E^{\mathrm{scat}}_k + E^{\mathrm{ann}}_k = \xi_0\gamma_0 m_\chi = \gamma_s m_\chi .
\label{eq:Ektotal}
\end{equation}

The annihilation factor depends on the relevant cross section and the effective volume occupied by DM particles after they thermalize in the center of the star, $A_\chi\sim \langle\sigma_{\mathrm{ann}}v\rangle/r^3_{0}$, where $r_{0} = \sqrt{3T_{\mathrm{DM}}/[2\pi G\,m_\chi(\rho_c+3p_c)]}$ and $p_c$ is pressure in the star's core. At late times and for elastic DM interactions, the DM is expected to eventually thermalize with the NS temperature, $T_{\mathrm{NS}}$. However, this process can be prolonged for specific DM scenarios~\cite{Garani:2020wge,Bell:2023ysh}; see also the discussion on the partial thermalization of inelastic DM~\cite{Acevedo:2024ttq}.

\begin{figure}
    \centering
    \includegraphics[width=0.7\textwidth]{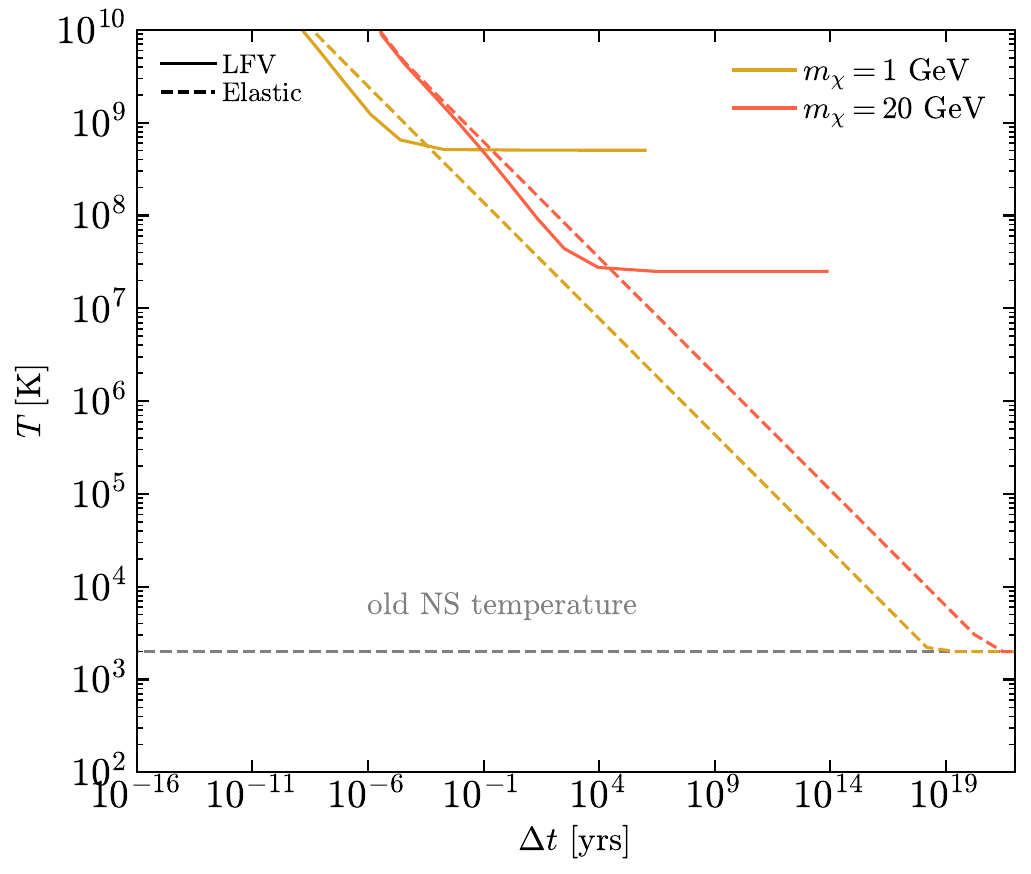}
    \caption{Evolution of the DM temperature ($T_{\mathrm{DM}}$) as a function of the time elapsed since capture, $\Delta t$, in years for $m_\chi = 2 \ m_a$ and $m_\chi = 1~\mathrm{GeV}$ (brown lines) or $20~\mathrm{GeV}$ (red). Solid lines correspond to LFV DM, while dashed lines represent the lepton flavor conserving case. The horizontal gray dashed line indicates the temperature of an old neutron star, $T_\mathrm{NS}$.
     \label{fig:t_vs_T}}
\end{figure}

 Here we point out that, when only LFV interactions are possible, the kinematics of inelastic scatterings with electron and muon targets in the star prevents DM from cooling below a threshold temperature. In \Cref{sec:t_th}, we show that for sufficiently small DM kinetic energies, the minimum target lepton energy that renders LFV interactions possible is given by $E_{p_1}^+ = \Delta m^2_{\mu e}/2q$, where $q$ is the three-momentum exchange and $\Delta m_{\mu e}^2 = m_\mu^2 - m_e^2$. This condition is most easily satisfied when  $q= q_{\textrm{max}} \simeq 2\sqrt{3m_\chi T_{\textrm{DM}}}$, which corresponds to a back-scattering event. We then find $E_{p_1}^+ = \Delta m^2_{\mu e}/4\sqrt{3m_\chi T_{\textrm{DM}}}$. As can be seen, as the DM temperature $T_{\textrm{DM}}$ decreases, the minimum required lepton energy $E_{p_1}^+$ can exceed the chemical potential close to the center of the star, $\mu\sim 250~\textrm{MeV}$, i.e. $E_{p_1}^+>\mu$. At this point, the LFV interactions can no longer thermalize DM particles within the star, leading to the following lower limit for the DM temperature
\begin{equation}
\begin{split}
T_\mathrm{DM} \gtrsim \frac{1}{3 m_\chi} \left( \frac{\Delta m_{\mu e}^2}{4 \mu} \right)^2\,.
\end{split}
\end{equation}
In \cref{fig:t_vs_T}, we illustrate the evolution of the DM temperature as a function of the time interval $\Delta t$ since the initial capture. In the plot, we compare this with the standard DM temperature evolution obtained when elastic (lepton flavor conserving) DM scattering off electron and muon targets is allowed. As can be seen, in the LFV case, DM rapidly equilibrates, but its temperature remains orders of magnitude higher than the NS temperature, $T_{\mathrm{DM}}\gg T_{\mathrm{NS}}$. Notably, for $m_\chi\sim \mathrm{GeV}$, this temperature, $T_{\rm DM, NS}\sim 43~\mathrm{keV}$, also exceeds the characteristic DM temperature in the Milky Way (MW), $T_{\rm DM,MW} \sim m_\chi v_\chi^2 \sim \textrm{keV}$ for $v_\chi\sim 10^{-3}$.

We note that, at very late times, DM is expected to fully thermalize with the star due to loop-induced processes, $\chi e\to \chi e$ and $\chi \mu\to \chi \mu$. These processes remain very strongly suppressed such that the relevant thermalization time significantly exceeds the age of the universe.

The increased DM temperature affects the interplay between the $p$-wave and $s$-wave annihilation modes inside the star because the annihilation factor varies differently with $T_{\mathrm{DM}}$ in both cases, $A_\chi\sim 1/T_{\mathrm{DM}}^{n/2}$, where $n = 3$ for $s$-wave and $n=1$ for $p$-wave. As a result, we find that for the allowed values of the $g_{\mu e}$ coupling and the values of the dark coupling constant $g_\chi$ corresponding to the thermal DM target, the annihilation heating of NSs for LFV interactions is driven by the $p$-wave process, $\chi\bar{\chi}\to aa$, due to the \textsl{flavor blocking} of DM thermalization in the star. 

The produced ALPs deposit their energy by decaying back into $e^{\pm}\mu^{\mp}$ pairs. For ALP masses close to the kinematical threshold, $m_a \sim m_\mu+m_e$, such decays remain Pauli-blocked in the NS inner core, where the lepton chemical potential is high. In this case, ALPs travel to the outer region of the star, where the chemical potential decreases, and deposit their energy there, therefore still heating the star.

\section{Results}

\begin{figure}
    \centering
    \includegraphics[width=0.9\textwidth]{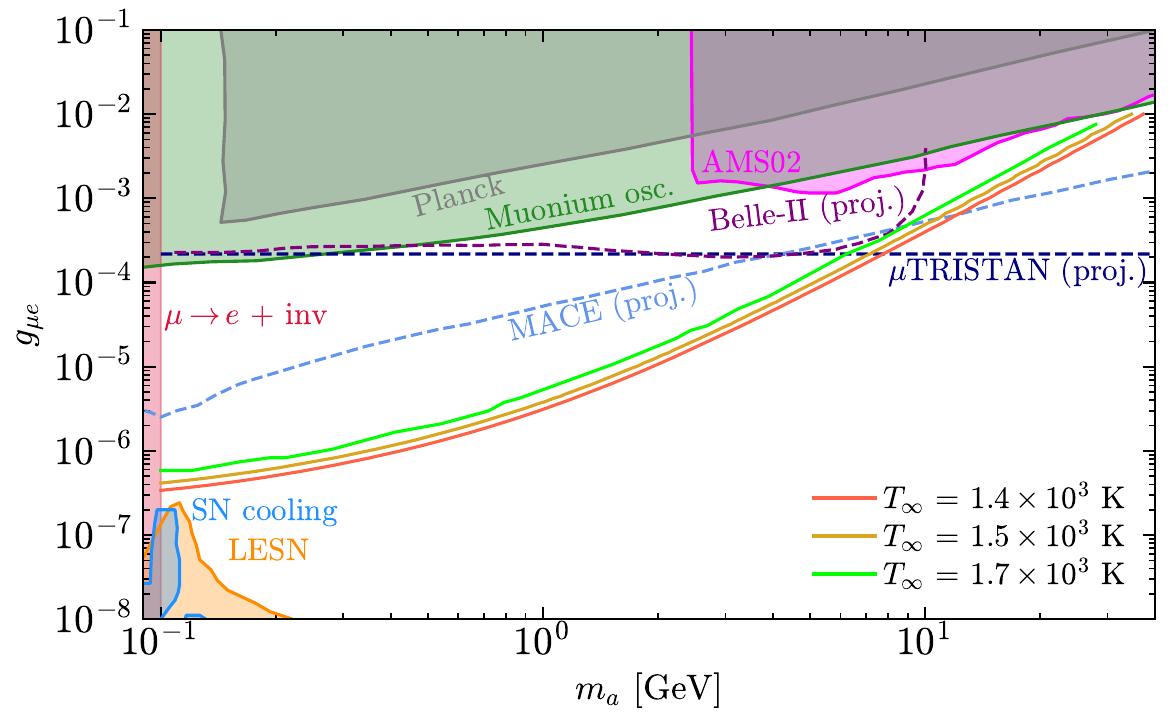}
    \caption{Current constraints and projected exclusion bounds on the LFV DM model with the ALP mediator and the $m_\chi = 2m_a$ mass hierarchy in the $(m_a,g_{\mu e})$ plane. The green (brown, red) solid lines define the lower boundary of the regions in the parameter space where DM-induced kinetic and annihilation heating is expected to sustain neutron star temperatures $T_\infty$ above $1.7\times 10^3~\mathrm{K}$ ($1.5\times 10^3~\mathrm{K}$, $1.4\times 10^3~\mathrm{K}$) even for old NSs.
     \label{fig:temp_lines_BSk25_case_1}}
\end{figure}

In \cref{fig:temp_lines_BSk25_case_1}, we illustrate the projected exclusion bounds on the LFV DM scenario that would result from future observations of NSs with temperatures $T_\infty\lesssim 1.7\times 10^3~\textrm{K}$. For sufficiently high $g_{\mu e}$, DM-induced heating is expected to prevent NSs from cooling below the temperatures indicated in the plot. The results shown here are obtained for $m_\chi = 2m_a$. For each DM mass, we assume the thermal value of the dark coupling constant $g_\chi$, which guarantees the correct DM relic density for any value of $g_{\mu e}$.

Current bounds on this scenario primarily result from searches for LFV ALPs. Light ALPs with $m_a< m_\mu - m_e$ are strongly constrained by the search for the invisible muon decay $\mu\to e+\textrm{inv.}$~\cite{Calibbi:2020jvd}. These particles, produced in the two-body muon decay, $\mu\to e a$, subsequently decay into electrons and neutrinos (e.g., $a\to e^+e^-\nu_\mu\bar{\nu}_e$) with a sufficiently large lifetime that they mimic invisible energy loss in the detector. For light DM, three-body decays via off-shell ALPs, $\mu\to e\chi\bar{\chi}$, are also possible. Light ALPs are, therefore, excluded in the $g_{\mu e}$ coupling regime of our interest. For small such couplings, astrophysical bounds derived from supernovae cooling and low-energy supernovae (LESN) explosions extend these constraints up to $m_a\sim 500~\textrm{MeV}$~\cite{Calibbi:2020jvd,Zhang:2023vva,Li:2025beu,Huang:2025rmy,Huang:2025xvo}. 

For even heavier ALPs, the upper bound on $g_{\mu e}\lesssim 10^{-3}$ is primarily driven by constraints from the search for spontaneous muonium-antimuonium conversion~\cite{Willmann:1998gd}; cf. also Refs~\cite{Dev:2017ftk,Calibbi:2024rcm}. In the plot, we also present relevant future projected constraints from the MACE experiment~\cite{Bai:2022sxq}. We also show the expected bounds from the search for LFV ALPs in the proposed $\mu$TRISTAN experiment~\cite{Calibbi:2024rcm} and the projected Belle-II sensitivity obtained with $50~\textrm{ab}^{-1}$ of integrated luminosity, derived from the search for explicit LFV in the $e^+e^-\to \mu^\pm\mu^\pm e^\mp e^\mp$ process~\cite{Endo:2020mev}. By contrast, the FCC-ee operating at the $Z$ pole is expected to achieve an even higher integrated luminosity of $150~\textrm{ab}^{-1}$~\cite{FCC:2018evy}. This could enable it to set yet more stringent bounds on $g_{\mu e}$ while extending the search toward higher ALP masses~\cite{Dev:2017ftk}. As can be seen, however, future NS bounds can surpass all these experimental projections for ALP masses up to $m_a\sim \textrm{a few GeV}$ and close the gap between them and supernovae constraints for $m_a\sim m_\mu$.

In the plot, we also present DM ID constraints derived from observations of the Cosmic Microwave Background (CMB) by the Planck collaboration~\cite{Planck:2018vyg} and from positron measurements by AMS-02~\cite{John:2021ugy}. Notably, these limits are based on the suppressed $s$-wave annihilation mode, $\chi\bar{\chi}\to e\mu$, which is sensitive to small values of $g_{\mu e}$. We rescale the more stringent existing bounds on DM annihilation into an $e^+e^-$ final state to account for the weaker impact of muons in the final state, cf. also recent analysis in Ref.~\cite{Liang:2025mfk}. As can be seen in the plot, the DM ID limits remain at most comparable to the muonium oscillation bound. A potential future improvement of these bounds by an order of magnitude in the annihilation cross section~\cite{Cooley:2022ufh,Wang:2025tdx} would help constrain $g_{\mu e}$ by an additional factor of a few.

Instead, future observations of old NSs could provide significantly stronger constraints on this scenario, potentially by three orders of magnitude in $g_{\mu e}$ at sub-GeV masses. Remarkably, the DM-induced NS heating is primarily driven by $p$-wave secluded annihilations, $\chi\bar{\chi}\to aa$, with the subsequent ALP decay inside the star, $a\to\mu e$. These $p$-wave annihilations are much less suppressed than for DM ID due to two key effects: \textsl{i)} boosting of DM particles in the NS gravitational field, and \textsl{ii)} flavor blocking of their equilibration with the star's temperature, which enhances the characteristic DM velocity in the NS. Ultimately, these results highlight that NS heating provides a window into this elusive dark sector, probing LFV thermal targets in regions of the parameter space that are otherwise quite challenging to access.

\section{Conclusions}

In this study, we have demonstrated that neutron stars can provide an interesting probe of dark sectors that are otherwise nearly invisible to traditional searches. Lepton flavor violating dark matter represents a particularly elusive thermal target; it evades direct detection through a lack of quark couplings, while its signature in electron scattering is suppressed by kinematic thresholds that non-relativistic halo DM cannot overcome. Furthermore, in the case of a pseudoscalar ALP mediator, indirect detection signals in the galactic halo are hampered by $p$-wave suppression at low velocities.

Our analysis highlights that the environment of a neutron star could act as a ``flavor laboratory'' that overcomes these barriers in several ways: \textsl{i)} the intense gravitational field accelerates DM to semi-relativistic speeds, providing the energy necessary for $e\to \mu$ upscattering that is kinematically forbidden on Earth; \textsl{ii)} the dense stellar interior provides access to muon targets, enabling $\mu\to e$ transitions; \textsl{iii)} the inelastic nature of the LFV interactions prevents the captured DM from reaching full thermal equilibrium with the star through a flavor-blocking mechanism. This keeps the DM population ``warm,'' which in turn prevents the $p$-wave annihilation cross-section from dropping to the negligible levels typically expected in cold astrophysical objects. Future extensions of this work could explore the impact of flavor-blocking on asymmetric or bosonic DM, where the altered isothermal distribution might lead to distinct signatures in stellar collapse~\cite{Zurek:2013wia}.

In the LFV DM case, these effects may allow future infrared observatories to probe couplings that are up to a few orders of magnitude smaller than what can be reached by the next generation of lepton colliders.  However, we note that such envisioned astrophysical measurements would require reaching well below current bounds \cite{Guillot:2019ugf} on NS temperatures, likely through dedicated observing campaigns. Beyond providing exclusion bounds, observations of old, isolated NSs could provide a novel signature for such dark sectors. While standard heating mechanisms, such as rotochemical heating~\cite{Fernandez:2005cg,Hamaguchi:2019oev} or superfluid vortex creep~\cite{Fujiwara:2023tmr,Fujiwara:2023hlj}, cf. review~\cite{Bramante:2023djs}, are intrinsically linked to the star’s magnetic and spin evolution, DM-induced heating creates a uniform temperature floor across the oldest stars in our galaxy. Observing a population of such stars with a uniform thermal profile could thus provide a compelling signature of dark matter, distinguishing it from standard late-stage stellar evolution.

In this context, identifying the LFV nature of the dark matter would rely on a synergy between diverse experimental frontiers: the absence of signals in traditional direct and indirect detection searches would reinforce the elusive profile of the dark sector, while potential hints of LFV at future leptonic colliders could provide the necessary evidence for the mediators responsible for these flavor-violating transitions.

Ultimately, this work bridges the gap between particle physics and astrophysics. If a LFV signal were to be detected in future accelerator facilities, neutron stars might provide further evidence that such a dark sector is responsible for the observed thermal relic density of the universe.

\vspace{0.5cm}

{\tt Digital data for this work, including the complete neutron star profile, can be found as  ancillary files in the arXiv submission~\cite{data}.}

\acknowledgments
We thank Leszek Zdunik for valuable discussions regarding the neutron star capture rate and for reading the manuscript. The work of H.~D. is supported by the US Department of Energy under Grant Contract DE-SC0012704.  J.~H.~Z. and S.~T. are supported by the National Science Centre, Poland (research grant No. 2021/42/E/ST2/00031). ST is also partially supported by Teaming for Excellence grant Astrocent Plus (GA: 101137080) funded by the European Union, with complementary national funding from the MNiSW (MNiSW/2025/DIR/811).

\vspace{0.5cm}

\appendix
\section{Capture rate of dark matter}
\label{sec:capture_rate}

\subsection{Neutron stars}

Neutron stars (NSs) are extremely dense stellar remnants. As schematically illustrated in \cref{fig:opticalfactor}, the NS crust typically extends to a depth of up to a few hundred meters from its surface. Below this, one finds the NS core with a high density, $\rho \gtrsim 1.4 \times 10^{14} \ \mathrm{g} \ \mathrm{cm}^{-3}$~\cite{Pearson:2018tkr}. In our analysis, we consider the standard case of $npe\mu$ matter, meaning we assume NSs are primarily composed of neutrons, protons, electrons, and muons, and we neglect the presence of hyperons. Many approaches exist to model the equation of state of an NS~\cite{Akmal:1998cf, Rikovska-Stone:2006gml, Goriely:2010bm, Pearson:2011zz, Pearson:2012hz, Goriely:2013xba, Potekhin:2013qqa, Kojo:2014rca, Baym:2017whm, Pearson:2018tkr, Annala:2019eax}. Among them, e.g., are the unified EoS for cold, non-accreting matter, based on Brussels-Montreal functionals~\cite{Goriely:2010bm, Pearson:2011zz, Pearson:2012hz, Goriely:2013xba, Potekhin:2013qqa, Pearson:2018tkr}, and the relativistic mean-field (RMF) approach~\cite{Chen:2014sca, Huang:2023grj}. We focus on old and cold stars and, for illustration, we employ the specific BSk-25 EoS (Brussels-Montreal). We use this EoS as an input to solve the structure equations for static matter, known as the Tolman-Oppenheimer-Volkoff (TOV) equations~\cite{Tolman:1939jz, Oppenheimer:1939ne}. 

In order to produce the NS profiles, we have used a version of the software \textit{CompactObject}~\cite{Huang:2023grj, Huang:2024ewv, Huang:2024rfg, Huang:2024rvj}, modified by us to print the radius, mass, pressure and energy density in each step of the numerical solution of the TOV equations. We also solve the equations for a range of energy densities at $r=0$, so that we can interpolate the density needed to produce a precise mass NS we want to generate. After generating the first elements of the profile, we use the EoS to obtain the neutron, proton and lepton chemical potentials, and the baryon number density by interpolating the expected values given a fixed energy density. For the number densities, we obtain the leptonic ones by considering a zero-temperature, free, relativistic Fermi gas:
\begin{equation}
n_l = \frac{\left( \mu_l^2 - m_l^2 \right)^{3/2}}{3 \pi^2} \times \Theta\left( \mu_l - m_l \right)\,,
\end{equation}
where $\mu_l$ and $m_l$ are the lepton chemical potential and masses respectively. This relation is valid for low temperatures as in our case. The number density of the protons is obtained by assuming charge neutrality: $n_p = n_e + n_\mu$. For the neutrons we use: $n_n = n_b - n_p$, where $n_b$ is the baryon number density. Finally, we get the elements of the metric by solving the following:
\begin{equation}
\begin{split}
|g_{tt}| &= \left( 1 - \frac{2 G M_\mathrm{NS}}{c^2 r} \right) \times \exp\left( -2 \int_{0}^{p} \frac{dp'}{p' + e(p')} \right)\,,\\
|g_{rr}| &= \left(1 - \frac{2Gm(r)}{c^2 r} \right)^{-1}\,,
\end{split}
\end{equation}
where $e(p)$ is the energy density as a function of pressure and the results obtained depend on the position ($r$) and the pressure at that point (for $g_{tt}$) and on the position and the mass inside a sphere of radius $r$ ($m(r)$). All this data is put into tables so that any needed quantity can be interpolated for a given $r$.

\subsection{Capture rate expression}

Capture of DM particles takes place as they are attracted by the gravitational force of the star and when they interact with its matter, so that the outgoing DM fermion cannot escape. In our case, since the interactions are with degenerate leptons and because we are considering very cold stars, if an interaction takes place, then the DM particle is automatically captured. This is because the target lepton has an energy below the leptonic chemical potential, $\mu_l$, while the outgoing above $\mu_l$. Therefore, the DM fermion will lose energy and, due to starting its motion roughly at rest, $v_\infty \sim 0$, then any energy loss will imply being gravitationally captured.

There are two possible interactions at tree level that can lead to capture DM: $\chi + e \to \chi + \mu$ and $\chi + \mu \to \chi + e$. We will work on the kinematics using the following notation: any 4-vector will just be represented by a letter with a subscript number ($P_i$), where $i=1$ will denote ingoing and $i=2$ outgoing. $k$ will be used for the DM particles, while $p$ for the leptons. The energy and 3-momentum magnitude of a particle with 4-momentum $P_i$ will be $E_{P_i}$ and $p_{P_i}$ respectively. 

The mean amplitude squared of any of the mentioned processes is:
\begin{equation}
\label{eq:SqAmp}
\begin{split}
\langle \left|\mathcal{M}\right|^2\rangle = \frac{g_\chi^2 g_{\mu e}^2}{(t - m_a^2)^2} t \ \ \times \left(t - m_\mu^2 + 2m_\mu m_e \cos \left( 2 \phi\right) - m_e^2 \right) \,.
\end{split}
\end{equation}
where $t \equiv \left( k_1 - k_2 \right)^2$ is the Mandelstam variable. The relevant differential cross section reads:
\begin{eqnarray}
\label{eq:d_sigma}
\frac{d\sigma}{d \cos{\theta}_\mathrm{cm}}_{\alpha \to \beta} = \frac{\langle\left|\mathcal{M}\right|^2\rangle}{32 \pi s} \frac{g_s (m_\beta)}{g_s (m_\alpha)}\,,
\end{eqnarray}
where $\theta_\mathrm{cm}$ is the angle between the incoming and outgoing DM particles in the CM frame, the target and outgoing leptons are $\ell_\alpha$ and $\ell_\beta$ respectively, $g_s (m) \equiv \sqrt{s^2 - 2 (m_\chi^2 + m^2) s + (m_\chi^2 - m^2)^2 }$ and $s \equiv \left(k_1 + p_1 \right)^2$ is the Mandelstam variable. 

Since we have two channels, the expression for the interaction rate takes a more complicated form than the one in Ref.~\cite{Bell:2020jou}: $\Omega^- \equiv \Omega^-_{\mu \to e} + \Omega^-_{e \to \mu}$, such that
\begin{equation}
\label{eq:omega_minus}
\begin{split}
\Omega^-_{\alpha \to \beta} = &\frac{\zeta}{32 \pi^3 m_\chi} \sqrt{\frac{g_{tt}}{1 - g_{tt}}}  \int_{m_\alpha}^{\mu_l} d E_{p_1} E_{p_1}  \int_{s_\mathrm{min}}^{s_\mathrm{max}} \frac{s ds}{g_s (m_\alpha) \beta_s (m_\alpha)} \int_{t_\mathrm{min}}^{t_\mathrm{max}} dt \langle\left|\mathcal{M} \right| \rangle \Theta (E_{p_2} - \mu_l) \,,
\end{split}
\end{equation}
where $\zeta \equiv n_\alpha / n_\mathrm{free}$ is a correction to the target number density as was computed by \cite{Garani:2018kkd}, $\mu_l$ is the leptonic chemical potential, $t \equiv \left(k_1 - k_2 \right)^2$ is the Mandelstam variable, $\beta_s (m) \equiv s^2 - (m_\chi^2 - m_\alpha^2)^2$. The Mandelstam variables limiting values are:
\begin{equation}
\begin{split}
 s_{\mathrm{min,1}} \le s \le s_{\mathrm{max}} \wedge (m_\chi + \mathrm{max} (m_{\alpha}, m_{\beta}))^2 \le s\,,
\end{split}
\end{equation}
such that $s_{\mathrm{min,1} / \mathrm{max}} \equiv m_\chi^2 + m_\alpha^2 + 2 E_{k_1} E_{p_1} \mp 2 p_{k_1} p_{p_1}$. $s_{\mathrm{min}}$ is the maximum value of $s_{\mathrm{min,1}}$ and $(m_\chi + \mathrm{max} (m_{\alpha}, m_{\beta}))^2$. For $t$ we have:
\begin{equation}
\begin{split}
t_{\mathrm{min} / \mathrm{max}} = &-\frac{s^2 - A_t s + B_t \pm \gamma_s\left( m_\alpha \right) \gamma_s\left( m_\beta \right)}{2 s} \,,
\end{split}
\end{equation}
where $A_t \equiv \left(m_{\alpha }^2+m_{\beta }^2+2 m_{\chi }^2\right)$ and $B_t = \left(m_{\alpha }^2-m_{\chi }^2\right) \left(m_{\beta }^2-m_{\chi }^2\right)$.
The energy of the outgoing target, $E_{p_2}$, is a long expression that depends on $E_{k_1}$, $E_{p_1}$, $s$, $t$ and the masses. This approach considers one scattering for capture. We have also considered multiple scattering~\cite{Bell:2020jou}, though they do not alter our results for the masses taken into account.

\subsection{Optical factor}

After defining the saturation limit as the minimum value of the cross section for which all DM particles get captured and the geometric limit as the minimum limit for which all the DM particles get captured within a thin shell close to the surface of the star, we can study the optical factor as a function of the radial position in the star for those limiting situation. 
The optical factor represents the amount of flux that remains along a trajectory for a point at $r$, given that the motion was from $r'=R_{\rm NS}$ to $r'=r$. To compute the relative quantity of flux that is absorbed for some radial position $r$, which is the only parameter needed due to spherical symmetry, we need to take into account the definition of the interaction rate, $\Omega^{-}(r)$, which 
is the probability rate of being captured as seen from far from the NS. Therefore, as seen locally, this quantity would be $\Omega^{-}(r)/\sqrt{g_{tt}(r)}$. In the following, we will leave the radial dependency implicit, for simplicity.

Taking the definition of $\Omega^{-}/\sqrt{g_{tt}}$, the equation for the evolution of the number of particles going through a particular geodesic at a position $r$ is:
\begin{equation}
\frac{dN}{dt^{\mathrm{loc}}}
    = -\frac{\Omega^{-}}{\sqrt{g_{tt}}}\, N,    
\end{equation}
whose solution is:
\begin{equation}
\begin{split}
\frac{N}{N_{0}}
    = e^{-\int dt^{\mathrm{loc}} \frac{\Omega^{-}}{\sqrt{g_{tt}}}}
    = \eta,
\end{split}
\end{equation}
where $\eta$ is the optical factor and represents the relative quantity of flux that reaches the position $r$ through a particular geodesic, and $t^{\mathrm{loc}}$ refers to local proper time in an orthonormal basis at $r$.

To compute kinematical quantities for a particular geodesic, we look at the equations of motion. These geodesic equations are written in terms of conserved quantities in the coordinate basis: $\mathscr{E} \equiv \frac{E_{\chi}^{\infty}}{m_{\chi}} = 1$ and $\mathscr{L} = \frac{J}{m_{\chi}}$. We replace these quantities directly and consider motion on the equatorial plane, $\theta = \pi/2$:
\begin{equation}
\begin{split}
\frac{dt}{d\tau} = &u^{t} = \frac{1}{g_{tt}}\,,\\
\frac{dr}{d\tau}
    = &u^{r}
    = \pm \sqrt{\frac{1}{g_{tt}g_{rr}}
        - \frac{1}{g_{rr}}\left(1 + \frac{J^{2}}{m_{\chi}^{2} r^{2}} \right)}\,,\\
\frac{d\phi}{d\tau} = &u^{\phi} = \frac{J}{m_{\chi} r^{2}}\,,\\
\frac{d\theta}{d\tau} = &u^{\theta} = 0\,.
\end{split}
\end{equation}
The sign of $u^{r}$ depends on whether the trajectory has already passed the minimum radius, $r = r_{\mathrm{min}}$. The coordinate time increment satisfies $dt^{\mathrm{loc}} = \sqrt{g_{tt}}\, dt$, so that,
\begin{equation}
\begin{split}
\eta = e^{-\int dt\,\Omega^{-}}\,.
\end{split}
\end{equation}
Since absorption depends only on $r$, we convert $dt$ to $dr$:
\begin{equation}
\begin{split}
\int dt\, \Omega^{-} &= \int dr'\, \frac{dt}{dr'}\, \Omega^{-} = \int dr'\, \frac{dt/d\tau}{dr'/d\tau}\, \Omega^{-} = \int dr'\,  \frac{1/g_{tt}}{\sqrt{\frac{1}{g_{tt}g_{rr}} - \frac{1}{g_{rr}}\left(1 + \frac{J^{2}}{m_{\chi}^{2} r'^{2}} \right)}}\, \Omega^{-}\,,\\
&= \int dr'\, \sqrt{\frac{g_{rr}}{g_{tt}(1-g_{tt})}} \frac{1}{\sqrt{1 - \frac{J^{2}}{\frac{1-g_{tt}}{g_{tt}}m_{\chi}^{2} r'^{2}}}}\, \Omega^{-}\,,
\end{split}
\end{equation}
where we use a prime, $r'$, not to confuse the radius along the geodesic with the point at which we are computing $\eta$ and all the metric elements are evaluated at $r'$.
This expression simplifies using the maximum angular momentum for which the radial motion is allowed:
\begin{equation}
\begin{split}
\frac{1}{g_{tt}g_{rr}}
    - \frac{1}{g_{rr}}
        \left( 1 + \frac{J^{2}}{m_{\chi}^{2}r^{2}} \right)
    \ge 0
\quad \Rightarrow \quad
J_{\max}^{2}
    = \frac{1-g_{tt}}{g_{tt}} m_{\chi}^{2} r^{2},
\end{split}
\end{equation}
so that:
\begin{equation}
\begin{split}
\int dt\,\Omega^{-}
    = \int dr'\, 
        \sqrt{\frac{g_{rr}}{g_{tt}(1-g_{tt})}}
        \frac{1}{\sqrt{1 - \frac{J^{2}}{J_{\max}^{2}(r')}}}
        \, \Omega^{-}\,,
\end{split}
\end{equation}
where we have found an extra factor, $\sqrt{g_{rr} / g_{tt}}$, with respect to previous treatments. We parameterize $J$ as a fraction $y$ of its maximum at $r$:
\[
J \equiv y\, J_{\max}(r),
\]
where $y = \sin\theta$ in classical terms, where $\theta$ is the angle between the geodesic and $\vec{r}$. Then:
\begin{equation}
\begin{split}
\int dt\,\Omega^{-}
    = \int dr'\, 
        \frac{\sqrt{g_{rr}}}{1 - g_{tt}}
        \frac{1}{\sqrt{
            1 - y^{2}\,
                \frac{J_{\max}^{2}(r)}{J_{\max}^{2}(r')}
        }}
        \, \Omega^{-}\,.
\end{split}
\end{equation}

For fixed $J$ (or $y$), there are two trajectories: one from the surface and just decreasing in $r'$ down to $r$, and another that goes to $r_{\min}$ and returns to $r$. These define two path integrals:
\begin{equation}
\label{eq:optical_depths}
\begin{split}
\tau_{\chi}^{-}(r,y)
    &= \int_{r}^{R_{\rm NS}} dr'\,
      \sqrt{\frac{g_{rr}}{g_{tt}(1-g_{tt})}}
      \frac{1}{\sqrt{
            1 - y^{2}\,
                \frac{J_{\max}^{2}(r)}{J_{\max}^{2}(r')}
      }}
      \, \Omega^{-}\,,\\
\tau_{\chi}^{+}(r,y)
    &= \tau_{\chi}^{-}(r,y)
      + 2 \int_{r_{\min}}^{r} dr'\,
      \sqrt{\frac{g_{rr}}{g_{tt}(1-g_{tt})}}
      \frac{1}{\sqrt{
            1 - y^{2}\,
                \frac{J_{\max}^{2}(r)}{J_{\max}^{2}(r')}
      }}
      \, \Omega^{-}\,.
\end{split}
\end{equation}

To obtain the total suppression factor, we average over the distribution of angular momenta:
\begin{equation}
\begin{split}
\eta
    = \frac{1}{2}
      \int f(y)\, dy\,
        \left[
            e^{-\tau_{\chi}^{-}(r,y)}
            + e^{-\tau_{\chi}^{+}(r,y)}
        \right]\,.
\end{split}
\end{equation}

Since $y = \sin\theta$ and $\theta$ is distributed as $p_{\theta}(\theta)\, d\theta = \sin\theta\, d\theta$, and using $\theta = \sin^{-1}y$ with $d\theta/dy = 1/\sqrt{1-y^{2}}$, we get:
\begin{equation}
p_{\theta}(\theta)\, d\theta
    = \frac{y}{\sqrt{1-y^{2}}}\, dy\,.
\end{equation}
Therefore, the final expression for $\eta$ is:
\begin{equation}
\eta = \frac{1}{2} \int \frac{y}{\sqrt{1-y^{2}}}\, dy
\left[
e^{-\tau_{\chi}^{-}(r,y)}
+ e^{-\tau_{\chi}^{+}(r,y)}
\right]\,.
\end{equation}

\subsection{Saturation and geometric limit}

We want to redefine and distinguish two concepts that were usually used interchangeably, but whose distinction is important to analyze the effects of capture rate: saturation and geometric limit. 

We define the saturation limit as the lower limit of the cross section (or of a coupling), for which the star starts to be able to capture all the DM particles in it. In terms of the optical factor, it would be the minimum value of the cross section (or couplings) for which $\eta (0) = 0$.

We will consider the geometric limit to be the lower limit of the cross section (or couplings) for which everything gets captured on a thin shell of negligible thickness, usually close to the NS's surface. In this case, the optical factor takes the form of a step function: $\eta (r) = \Theta(r - R_\mathrm{eff})$, where $R_\mathrm{eff}$ is the position of the shell. In general, this shell is the NS surface. However, in our case, since the outer part of the star just has electrons and since the upscattering to muons is just permitted for some limiting lower value of the chemical potential, so that the electrons have enough energy to upscatter to a muon, there is an effective radius that we have found to be greater than the muon radius (the NS region with muons), $R_\mu$, but less than the radius of the whole star: $R_\mu \le R_\mathrm{eff} \le R_\mathrm{NS}$.

For all the parameters space between these two situations, everything is captured and the capture rate in this regime is~\cite{Bell:2018pkk}:
\begin{equation}
\begin{split}
C_\mathrm{NS} = \frac{\pi R_\mathrm{NS}^2 (1 - B (R_\mathrm{NS}))}{v_\mathrm{NS} B (R_\mathrm{NS})} \frac{\rho_\chi}{m_\chi} \mathrm{Erf} \left(\sqrt{\frac{3}{2}} \frac{v_\mathrm{NS}}{v_d} \right)\,,
\end{split}
\end{equation}
where $B(r) \equiv 1 - 2 G M_\mathrm{NS} / c^2r$. As it will be seen in the results, $R_{\rm NS}$ should really be $R_\mathrm{eff}$, the effective radius where interactions take place when we go to the saturation limit. However, this radius is very close to the radius and the difference with respect to the whole star is just negligible.

\section{Kinetic energy deposition}
\label{sec:kin_dep}

Because of the subtleties of General Relativity and the non-conservation of local energies, it is important to compute the deposition from the first interaction until the DM loses almost all its kinetic energy. The injection of energy converges to a uniform value when the DM loses almost all of its kinetic energy independently of the number of interactions $N$. We will not make any assumptions about neither where the interactions take place nor how many there are. We will just suppose the interactions end up at the center of the star, such that their sequence is at arbitrary radial distances ($r_i$, where $0 \leq i \leq N$) from the center of the star. This sequence goes from the first point where it interacts to the center of the star, where it finishes its journey. We will measure all the energies as seen from the surface. We will not assume the particle has a radial trajectory.

After the DM particle has its $n$th interaction, it loses a fraction $p_n$ ($0 \leq p_n \leq 1$) of its kinetic energy and gains energy before interacting again from its gravitational pull to the center of the star. To compute the energy gained just by the interaction with the classical gravitational field, we need to consider the invariant quantity: $\mathcal{E} \equiv \left| g_{\mu \nu} p^\mu \xi^{\nu} \right|$, where $\xi^{\nu}$ is the Killing vector equal to $(1, 0, 0, 0)$. If every element of the metric is represented as a positive quantity, the metric around a NS has the form:
\begin{equation}
ds^2 = - g_{tt} \  dt^2 + g_{rr} \  dr^2 + r^2 d\Omega^2\,.
\end{equation}
Let us consider the invariance of $\mathcal{E}$ along two different points in the trajectory of a free-falling DM particle: at $r = r_a$ and $r = r_b$. In that case, we have: $\mathcal{E} = g_{tt}^{(a)} E_a = g_{tt}^{(b)} E_b$. We need the vierbein $e^{\hat{t}}_t \equiv \sqrt{g_{tt}}$ in order to relate these quantities with the local energy [denoted by superscript $(0)$] that is involved in the local interactions we are taking into account:
\begin{equation}
E_i^{(0)} = \sqrt{g_{tt}^{(i)}} E_i\,,
\end{equation}
where $g_{tt}^{(i)}$ is the metric $tt$ element evaluated at $r_i$. With these two equalities, we can finally relate the two local energies:
\begin{equation}
\sqrt{g_{tt}^{(a)}} E_a^{(0)} = \sqrt{g_{tt}^{(b)}} E_b^{(0)}\,.
\end{equation}

Let us call $E_k^{(N)}$ the kinetic energy deposited after $N + 1$ interactions (from $n=0$ to $n=N$) and $E_N$ the remaining energy of the particle after those interactions, both  as seen from the surface of the NS. We could follow an iterative process as depicted as follows. We first compute the total energy at the place of interaction, $E_n^\mathrm{tot}$, when the $n$th interaction takes place at the position $r_n$. For $n = 0$, it is given by the boost of the mass, $\gamma_{r_0}$, where $r_0$ is the position of the first interaction. With this energy we can compute the deposited kinetic energy, as seen from the surface, which is $K_n^\mathrm{in} = \xi_{r_n} p_n \left( E_n^\mathrm{tot} - m_\chi \right)$. The outgoing energy, in the reference frame of the $n$th interaction is  $K_n^\mathrm{out} = (1 - p_n) \left( E_n^\mathrm{tot} - m_\chi \right)$. With these two quantities we can compute the next total energy, as $E_{n + 1}^\mathrm{tot} = \frac{\gamma_{r_{n + 1}}}{\gamma_{r_{n}}} \left( K_n^\mathrm{out} + m_\chi  \right)$. We do this process and collect all the energy deposited into the star, which takes the form:
\begin{equation}
\begin{split}
E_k^{(N)} = \gamma_s m_\chi \sum_{n=0}^{N} p_n \left(1 - \frac{1}{\gamma_{r_n}} \right) \prod_{i=n+1}^{N+1} \left( 1 - p_i \right)\,,
\end{split}
\end{equation}
such that $p_{N+1} \equiv 0$ because we consider $N$ to be the last possible interaction before flavor blocking and $s$ stands for ``surface."  By computing the final energy left, we can see that it is:
\begin{equation}
\begin{split}
E_N = \gamma_s m_\chi - E_k^{(N)}\,.
\end{split}
\end{equation}
However, we claim that the DM particle has lost almost all of its energy and to be located close to the center of the star. Therefore, we expect this remaining energy to be $E_N = \xi_0 m_\chi$, where $0$ stands for $r = 0$, the center of the NS. So we can compute the total deposited energy as:
\begin{equation}
\begin{split}
E_k^{(N)} = \left(\gamma_s - \xi_0 \right) m_\chi\,.
\end{split}
\end{equation}
If we also consider annihilations at the center of the star and assuming that we have reached the equilibrium time, where the number of DM particles annihilating equals the number of the ones captured, then we have to add $E_N$ and the expression modifies to:
\begin{equation}
E_k^{(N)} = \gamma_s m_\chi\,,
\end{equation}
which is the total energy of the particle as it goes through the surface of the star. This quantity differs from the one in~\cite{Bell:2023ysh}, where the boost is considered at the center of the star and which may overestimate the deposited energy.

This approach establishes the energy used to compute the heating of the NS at the surface. The energy deposited by each particle in its first interaction as seen from the surface, i.e. $E_k \equiv \xi_r q_0 = \xi_r \left(E_{k_1} - E_{k_2}\right)$, would underestimate the heating. This energy has two main components, $q_0 = E_\mathrm{Auger} + \Delta E$: the Auger energy and the energy difference above the Fermi sea. The outgoing lepton must have an energy above the Fermi see, $E_{p_2} > \mu_l$, due to Pauli blocking. The space occupied by the target lepton is rapidly filled by nearby leptons below the Fermi sea, generating a rearrange of leptons that liberates an energy equal to the difference between the chemical potential and the target lepton: the Auger energy~\cite{Bramante:2023djs}. An extra contribution comes when the produced lepton is reabsorbed near the surface of the Fermi sea, depositing the excess of energy it had with respect to $\mu_l$: $\Delta E \equiv E_{p_2} - \mu_l$. There is still another contribution that may come from the different masses of the leptons, the target and the produced. However, to keep the thermal distributions in the star, the leptons need to reconvert so that the total number of each species remains constant. Therefore, this contribution is neglected.

\section{Interaction rate of LFV dark matter}
In order to compute the time required by the captured DM particles to thermalize in the center of the star, we need to compute the interaction rate, which follows from Fermi's Golden Rule:
\begin{equation}
\begin{split}
\Gamma = &2\int \mathcal{D}p_{p_1} \mathcal{D}p_{k_2} \mathcal{D}p_{p_2} \frac{\langle \left| \mathcal{M} \right|^2 \rangle} {2 E_{k_1}} \left(2 \pi \right)^4 \delta^{(4)}(k_1 + p_1 - k_2 - p_2) n_F\left(E_{p_1}\right) \left(1 - n_F\left(E_{p_2}\right)\right)\,,
\end{split}
\end{equation}
where $\mathcal{D}p \equiv \frac{d^3p}{(2\pi)^3 2 E_p}$ and $\langle \left| \mathcal{M} \right|^2 \rangle$ is the squared amplitude of the process, summed over all possible final spins and averaged over the initial ones, and it is equal to:
\begin{equation}
\begin{split}
\langle \left| \mathcal{M} \right|^2 \rangle = &\frac{A t + B t^2}{\left(t - m_a^2 \right)^2}\,,
\end{split}
\end{equation}
where $A \equiv - g_{\mu e}^2 g_\chi^2 (m_\mu^2 + m_e^2 - 2 m_\mu m_e \cos 2\phi)$ and $B \equiv g_{\mu e}^2 g_\chi^2$. Following \cite{Bell:2020jou}, we will express $\Gamma$ as:
\begin{equation}
\begin{split}
\Gamma = &\int \frac{d^3 p_{k_2}}{\left(2 \pi \right)^3} \frac{\langle \left| \mathcal{M} \right|^2 \rangle}{\left(2 E_{k_1}\right) \left(2 E_{k_2}\right) \left(2 m_{1}\right) \left(2 m_{2}\right)} \Theta\left( E_{k_2} - m_\chi \right) \Theta\left( q_0 \right) S\left(q_0, q \right)\,,
\end{split}
\end{equation}
where the first $\Theta-$ function guarantees that the outgoing DM particle has an energy greater than its mass, the second one forces the energy transfer to go from the DM current to the leptonic one (which is demanded by Pauli blocking at $T \to 0$), $S\left(q_0, q \right)$ is expressed in terms of $q_0$ and $q$, the time and spatial components of the transverse momentum, so that we can transform the integral in $p_{k_2}$ into those variables. $S\left(q_0, q \right)$ is explicitly:
\begin{equation}
\begin{split}
 S\left(q_0, q \right) = &8 m_1 m_2 \int \mathcal{D}p_{p_1} \mathcal{D}p_{p_2} \left(2 \pi \right)^4 \delta^{(4)}(k_1 + p_1 - k_2 - p_2) n_F\left(E_{p_1}\right) \left(1 - n_F\left(E_{p_1}\right)\right) \\ & \ \ \ \ \ \ \ \ \ \ \ \ \ \ \ \ \times \Theta\left( E_{p_1} - m_1 \right) \Theta\left( E_{p_2} - m_2 \right),\\
 = &\frac{m_1 m_2}{2\pi^2} \int \frac{d^3 p_{p_1}}{E_{p_1} E_{p_2}} \delta \left( q_0 + E_{p_1} - E_{p_2}\right) n_F\left(E_{p_1}\right) \left(1 - n_F\left(E_{p_1}\right)\right) \Theta\left( E_{p_1} - m_1 \right) \Theta\left( E_{p_2} - m_2 \right)\,.
\end{split}
\end{equation}
We have added again $\Theta-$functions to force the energies to higher values than the masses. After integrating on $p_{p_2}$ with the help of the $\delta-$distribution we can express $E_{p_2}$ in terms of $p_{p_1}$, $q$ and $\theta_{q1}$, the angle between $\vec{q}$ and $\vec{p}_1$:
\begin{equation}
\begin{split}
E_{p_2} &= \sqrt{m_2^2 + p_{p_1}^2 + q^2 + 2p_{p_1} q \cos \theta_{q1}}\\
&= \sqrt{E_{p_1}^2 + \Delta m_{21}^2 + q^2 + 2p_{p_1} q \cos \theta_{q1}} > m_2\,,
\end{split}
\end{equation}
where $\Delta m_{21}^2 \equiv m_2^2 - m_1^2$. That expression for $E_{p_2}$ goes in the Dirac $\delta-$distribution: $\delta \left( q_0 + E_{p_1} - E_{p_2}\right) \to \delta \left( q_0 + E_{p_1} - \sqrt{E_{p_1}^2 + \Delta m_{21}^2 + q^2 + 2p_{p_1} q \cos \theta_{q1}}\right)$. We can also change the variable $p_{p_1}$ to $E_{p_1}$ by taking into account that $p_{p_1} dp_{p_1} = E_{p_1} dE_{p_1}$, so that: $d^3 p_{p_1} = 2 \pi p_{p_1} E_{p_1} dE_{p_1} d\cos \theta_{q1}$, where we took the $\hat{z}$ direction pointing towards $\vec{q}$. 

Now, we can evaluate the Dirac $\delta$ by integrating it in $\cos \theta_{q1}$. There are two expressions we need:
\begin{enumerate}
    \item $\left|\frac{d}{d \cos \theta_{q1}}\left( q_0 + E_{p_1} - \sqrt{E_{p_1}^2 + \Delta m_{21}^2 + q^2 + 2p_{p_1} q \cos \theta_{q1}} \right) \right| = \frac{p_{p_1} q}{E_{p_2}}$,
    \item $q_0 + E_{p_1} - \sqrt{E_{p_1}^2 + \Delta m_{21}^2 + q^2 + 2p_{p_1} q \cos \theta_{q1}} = 0 \to \cos \theta_{q1} = \frac{q_0^2 - q^2 + 2 q_0 E_{p_1} - \Delta m_{21}^2}{2 p_{p_1} q}$.
\end{enumerate}

After replacing these quantities in the expression for $S \left(q_0, q \right)$, we obtain:
\begin{equation}
\begin{split}
 S\left(q_0, q \right) = &\frac{m_1 m_2}{\pi q} \int dE_{p_1} d\cos \theta_{q1} \delta \left( \cos \theta_{q1} - \frac{q_0^2 - q^2 + 2 q_0 E_{p_1} - \Delta m_{21}^2}{2 p_{p_1} q} \right) n_F\left(E_{p_1}\right) \left(1 - n_F\left(E_{p_1}\right)\right) \\ & \ \ \ \ \ \ \ \ \ \ \ \ \ \ \times \Theta\left( E_{p_1} - m_1 \right) \Theta\left( E_{p_2} - m_2 \right)\\
 = &\frac{m_1 m_2}{\pi q} \int_{m_1}^{\mu} dE_{p_1} \Theta\left( E_{p_2} - \mu \right)\,,
\end{split}
\end{equation}
where the Fermi distributions were  treated for $T \to 0$, which is a very good approximation for the masses involved, so that the lepton target must have an energy less than the chemical potential, $\mu$, and the outgoing lepton an energy greater than it.

The limits of integration for $E_{p_1}$ need further analysis. We will obtain two more constraints to those limits. First, from the Pauli blocking condition to the outgoing lepton we have: $E_{p_2} = E_{p_1} + q_0 \ge \mu$. Therefore, $E_{p_1} \ge \mu - q_0$. Second, from the condition that the absolute value of the expression found for $\cos \theta_{q1}$ is less or equal than $1$, we uncover another condition:
\begin{equation}
\nonumber
\begin{split}
&\cos^2 \theta_{q1} = \left(\frac{q_0^2 - q^2 + 2 q_0 E_{p_1} - \Delta m_{21}^2}{2 p_{p_1} q}\right)^2 \le 1\,.\\
&\text{Therefore:}\\
&4 t E_{p_1}^2 + 4 q_0 \left(t-\text{$\Delta $m}_{21}^2\right) E_{p_1} + 4 m_1^2 q^2+\left(t-\text{$\Delta $m}_{21}^2\right)^2 \leq 0\,,\\
&\text{whose roots are:}\\
&E_{p_1}^\mp = -\frac{q_0 \left(t-\text{$\Delta $m}_{21}^2\right)}{2t} \pm \frac{q \sqrt{\left(t-\text{$\Delta $m}_{21}^2\right)^2 - 4 m_1^2 t}}{2 t}\,.
\end{split}
\end{equation}
Taking into account that $t < 0$, the condition for $E_{p_1}$ is:
\begin{equation}
\label{eq:cond_Ep1}
\begin{split}
E_{p_1} \le E_{p_1}^- \lor  E_{p_1}^+ \le E_{p_1}\,.
\end{split}
\end{equation}
We need two conditions to be met: (1) $\max\left(m_1, \mu - q_0 \right) \le E_{p_1} \le  \mu$ and (2) \cref{eq:cond_Ep1}, which will be called region $\mathcal{C}$. If $I_E \left(q_0, q \right)$ is the length of this interval, we have the final expression for $S \left( q_0, q \right)$:
\begin{equation}
\begin{split}
 S\left(q_0, q \right) = &\frac{m_1 m_2}{\pi q} \int_{\mathcal{C}} dE_{p_1} \Theta\left( E_{p_2} - \mu \right)\\
 = &\frac{m_1 m_2}{\pi q} I_E \left(q_0, q \right) \Theta\left( E_{p_2} - \mu \right)\,.
\end{split}
\end{equation}
Depending on the point in phase space, the value of $I_E \left(q_0, q \right)$ is $\left[\mu - \max\left(m_1, \mu - q_0, E_{p_1}^+ \right) \right] \times \Theta\left(\mu -  E_{p_1}^+\right) + \left[\min\left(E_{p_1}^-, \mu \right) - \max\left(m_1, \mu - q_0 \right) \right] \times \Theta\left(E_{p_1}^- -  \max\left(m_1, \mu - q_0 \right)\right)$.

Now we need to transform the integral in $p_{k_2}$ into $q_0$ and $q$. First, we will express the expression as follows:
\begin{equation}
\begin{split}
\Gamma = \frac{1}{64 \pi^2 m_1 m_2} \int \frac{p_{k_2}^2 \ dp_{k_2} \ d\cos \theta_{k_1 k_2}}{E_{k_1} E_{k_2}} \frac{A t + B t^2}{\left(t - m_a^2 \right)^2} \Theta \left( E_{k_2} - m_\chi \right) \Theta \left( q_0 \right) S\left( q_0, q \right)\,,
\end{split}
\end{equation}
where $\theta_{k_1 k_2}$ is the angle between $\vec{k}_2$ and $\vec{k}_1$. The relations between the variables are the following:
\begin{equation}
\nonumber
\begin{split}
p_{k_2} = \sqrt{\left( E_{k_1} - q_0 \right)^2 - m_\chi^2} &\to dp_{k_2} = -\frac{\left( E_{k_1} - q_0 \right)}{p_{k_2}} dq_0\,,\\
\cos \theta_{k_1 k_2} = \frac{p_{k_1}^2 + p_{k_2}^2 - q^2}{2 p_{k_1} p_{k_2}} &\to d\cos \theta_{k_1 k_2} = -\frac{q}{p_{k_1} p_{k_2}} dq\,.
\end{split}
\end{equation}
With these quantities, we can transform the measure of the integral:
\begin{equation}
dp_{k_2} d\cos \theta_{k_1 k_2} = dq_0 dq \frac{\partial(p_{k_2}, \cos \theta_{k_1 k_2})}{\partial(q_0, q)} = dq_0 dq \frac{E_{k_2} q}{p_{k_1} p_{k_2}^2}\,.
\end{equation}
Then, the interaction rate takes the following form:
\begin{equation}
\begin{split}
\Gamma &= \frac{1}{64 \pi^2 m_1 m_2 E_{k_1} p_{k_1}} \int dq_0 dq q \frac{A t + B t^2}{\left(t - m_a^2 \right)^2} \Theta \left( E_{k_2} - m_\chi \right) \Theta \left( q_0 \right) S\left( q_0, q \right)\\
&= \frac{1}{64 \pi^3 E_{k_1} p_{k_1}} \int dq_0 dq \frac{A t + B t^2}{\left(t - m_a^2 \right)^2} I_E \left(q_0, q \right)\\ & \ \ \ \ \ \ \ \ \ \ \ \ \ \ \ \ \ \ \ \ \ \ \ \ \ \ \ \ \ \ \ \times \Theta \left( E_{k_2} - m_\chi \right) \Theta \left( q_0 \right)  \Theta\left( E_{p_2} - \mu \right)\,,
\end{split}
\end{equation}
Now we need to obtain the limits of integration:
\begin{equation}
\nonumber
\begin{split}
&\cos^2 \theta_{k_1 k_2} = \left(\frac{p_{k_1}^2 + \left(E_{k_1} - q_0\right)^2 - m_\chi^2 - q^2}{2 p_{k_1} \sqrt{\left( E_{k_1} - q_0 \right)^2 - m_\chi^2}}\right)^2 \le 1\,.\\
&\text{Therefore:}\\
&q^4 - 2 \left( \left(E_{k_1} - q_0 \right)^2 - m_\chi^2 + p_{k_1}^2 \right) q^2 + \left( \left(E_{k_1} - q_0 \right)^2 - m_\chi^2 - p_{k_1}^2 \right)^2 \leq 0\,,\\
&\text{whose roots are:}\\
&q^2_\pm = \left( \left( E_{k_1} - q_0 \right)^2 - m_\chi^2 + p_{k_1}^2  \right) \pm 2 p_{k_1} \sqrt{\left(E_{k_1} - q_0 \right)^2 - m_\chi^2}\,.
\end{split}
\end{equation}
It can be shown that the last expression is equivalent to: $q^2_\pm = \left( p_{k_1} \pm p_{k_2}  \right)^2$, therefore:
\begin{equation}
p_{k_1} - p_{k_2} \le q \le p_{k_1} + p_{k_2}\,,
\end{equation}
such that $p_{k_2} = \sqrt{\left(E_{k_1} - q_0 \right)^2- m_\chi^2}$. In addition, in the case of $q_0$, the limiting values are transferring no energy or all the kinetic energy from the incoming DM particle to the leptonic current. So the final expression for the interaction rate is:
\begin{equation}
\begin{split}
\Gamma &= \frac{1}{64 \pi^3 E_{k_1} p_{k_1}} \int_0^{E_{k_1} - m_\chi} dq_0 \int_{p_{k_1} - \sqrt{\left(E_{k_1} - q_0 \right) - m_\chi^2} }^{p_{k_1} + \sqrt{\left(E_{k_1} - q_0 \right) - m_\chi^2}} dq \frac{A t + B t^2}{\left(t - m_a^2 \right)^2} I_E \left(q_0, q \right)\,.
\end{split}
\end{equation}
In this expression, $t$ needs to be expressed in terms of the integration variables: $t = q_0^2 - q^2$. It is also important to note that, since $q_0 \ge 0$ and all the DM particles have the same mass, $p_{k_1} \ge p_{k_2}$ and it is not necessary to impose the absolute value of $p_{k_1} - p_{k_2}$ in the second integral above. Given the presence of many channels for capturing the DM, we can compute the total interaction rate as the sum of the rate for each channel, $\Gamma_i$:
\begin{equation}
\Gamma_\mathrm{tot} \equiv \sum_i \Gamma_i\,.
\end{equation}

\section{Thermalization and annihilation of LFV dark matter}

\subsection{Thermalization time}
\label{sec:t_th}

\subsubsection{General treatment}

To find the thermalization time, we need to find the time for each interaction until the mean kinetic energy is equal to $E_f - m_\chi \equiv \frac{3}{2} T_{NS}$, where $T_{NS}$ is the temperature of the star and also the DM in it. The time for each interaction is computed as the inverse of the interaction rate, found in the previous section. The process of computation consists in finding the mean energy lost by the DM particles in each interaction, so that the overall energy can be updated until the kinetic energy reaches its final estimated value, $E_f - m_\chi$. The initial energy is $\gamma m_\chi$.

The mean kinetic energy transferred in an interaction is just the average of $q_0$ weighted by the interaction rate itself~\cite{Bell:2020jou}:
\begin{equation}
\langle K \rangle = \frac{1}{\Gamma} \int dq_0 q_0 \frac{d \Gamma}{dq_0}\,.
\end{equation}
In our case, in which there is more than one channel of interaction, we compute the average weighted over the interaction rate for each channel:
\begin{equation}
\langle K \rangle = \sum_i \frac{\Gamma_i \times \langle K \rangle_i}{\Gamma_\mathrm{tot}} \,,
\end{equation}
where $\langle K \rangle_i$ is the mean kinetic energy transferred computed just for one channel. 
Therefore, for the $i$-th interaction of a DM particle of energy $E_i$, the outgoing energy will on average be equal to $E_{i+1} = E_i - \langle K \rangle$. Then, the thermalization time is:
\begin{equation}
\tau_\mathrm{th} = \sum_{i=1}^N \frac{1}{\Gamma(E_i)}\,,
\end{equation}
such that $E_N \le E_f < E_{N-1}$, where $E_f$ is the final energy attained by the iterative process.

\subsubsection{Inelastic processes subtleties}

For usual elastic processes, the procedure described just above can be performed down to any temperature, especially the temperature of the star, which is the relevant one to speak about thermalization. In our case, we have an inelastic process. When the DM particles have lost most of their energy, the process $\chi + l_\alpha \to \chi + l_\beta$ is simply forbidden when $m_\alpha \neq m_\beta$. An intuitive way to see this is the following: when the kinetic energy is low enough, we may regard the incoming and outgoing DM particles nearly at rest. Then, on the leptonic side of the process, we need a target lepton with an energy $E_{p_1} \lesssim \mu$, while the outgoing energy is $E_{p_2} \gtrsim \mu$. If we impose energy conservation, we cannot achieve momentum conservation, since both particles have different mass. 

\begin{figure}
    \centering
    \includegraphics[width=0.65\textwidth]{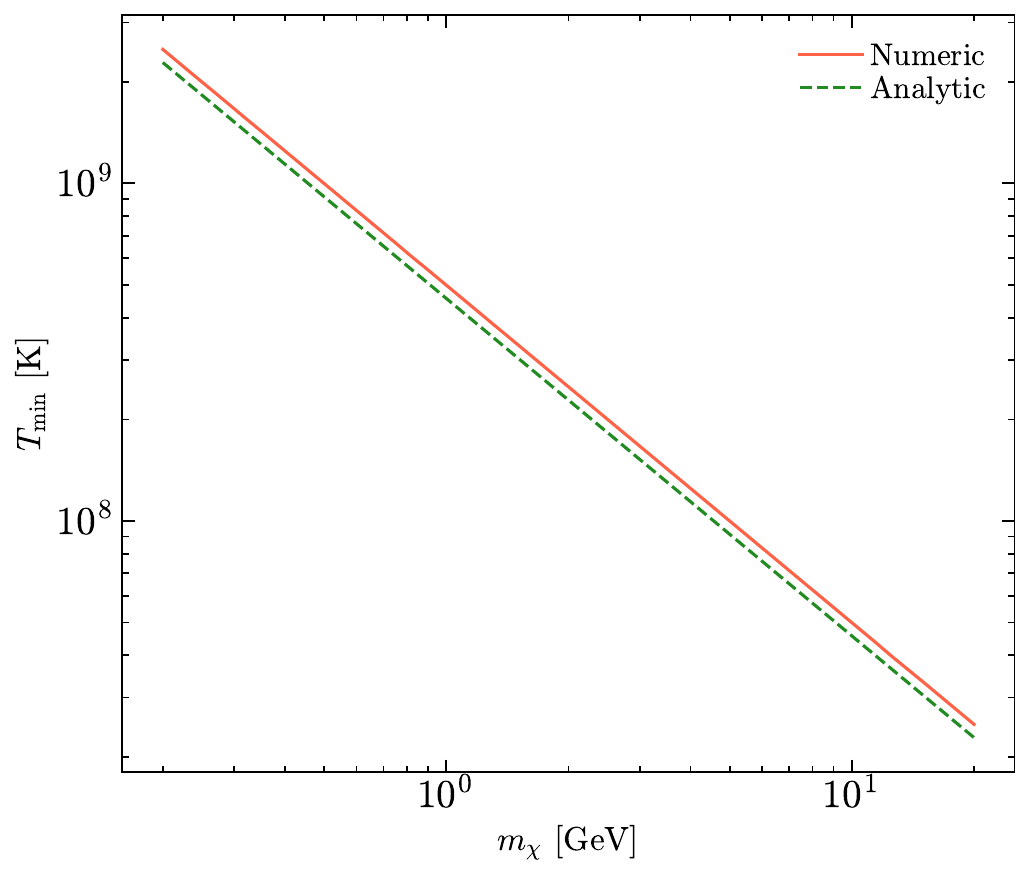}
    \caption{ Lowest temperature reached by DM for case $1$ and a NS of mass of $2 \ M_\odot$ (BSk 25).
     \label{fig:min_T}}
\end{figure}

We can also show this in a more rigorous way. The roots in \Cref{eq:cond_Ep1} can be expanded in powers of $q$ and $q_0$. We are considering them to be very small, since we are studying the behavior of the conditions for very low temperatures. Their values to first order are
\begin{equation}
\begin{split}
E_{p_1}^\pm = \pm \frac{\left| \Delta m_{\mu e}^2 \right|}{2 q} + \mathcal{O}\left( q^0, q_0^0 \right)\,,
\end{split}
\end{equation}
where $\mathcal{O}\left( q^0, q_0^0 \right)$ means that we are just keeping the poles in a Laurent series expansion. The first term is the one that dominates for small $q$ and $q_0$. It is clear that $E_{p_1}^-$ is in this case negative and by setting $E_{p_1}^+ > \mu$, we can be sure that there is no available target lepton with that energy. In this way, we freeze the transfer of energy between the DM and the NS. The minimum temperature is attained when $E_{p_1}^+ \simeq \frac{\left| \Delta m_{\mu e}^2 \right|}{2 q} \sim \mu$, therefore, when $q \sim \frac{\left| \Delta m_{\mu e}^2 \right|}{2 \mu}$. As can be seen, the interaction is totally excluded when the maximum value of $q$, $q_\mathrm{max} = p_{k_1} + p_{k_2}$, is excluded. To obtain this limit, we need to consider that the kinetic energy left, $K$, is related to the temperature: $K = \frac{3}{2} T$. So, if the temperature is really smaller than the mass of the DM, we can approximate $p_{k_1} = \sqrt{3m_\chi T} \simeq p_{k_2}$. Here we are assuming the value of $q_0$ is negligible, which is actually the case for our Laurent expansion of $E_{p_1}^+$. Therefore, we can express q in terms of the temperature and put it in the limit when any interaction is excluded at three level: $q_\mathrm{max} = 2\sqrt{3m_\chi T} \sim \frac{\left| \Delta m_{\mu e}^2 \right|}{2 \mu}$. By applying this approximation, we get a very similar result than performing the whole computation until the temperature DM stabilizes and ceases to decrease, as seen in \Cref{fig:min_T}. The minimum value of the temperature is:
\begin{equation}
\begin{split}
T_\mathrm{min} \simeq \frac{1}{3 m_\chi} \left( \frac{\Delta m_{\mu e}^2}{4 \mu} \right)^2\,.
\end{split}
\end{equation}

\subsubsection{Annihilation of DM in the star}

Although the DM accumulated in the center of the star cannot reach the temperature of the star, it can still annihilate and contribute in this way to heating the NS. The amount of DM fermions annihilating per time is~\cite{Bell:2023ysh}:
\begin{equation}
\begin{split}
\frac{dN_\chi^\mathrm{ann}}{dt} = A_\chi N_\chi^2\,,
\end{split}
\end{equation}
where $A_\chi$ is an annihilation factor that appears in the annihilation rate of DM: $\Gamma_\mathrm{ann} = \frac{1}{2} A N_\chi^2$. It can be approximated as~\cite{Bell:2023ysh}:
\begin{equation}
\begin{split}
A_\chi = \frac{\langle \sigma_\mathrm{ann} v_\chi \rangle}{N_\chi^2} \int n_\chi^2 (r) d^3r \sim \frac{\langle \sigma_\mathrm{ann} v_\chi \rangle}{\left( 2\pi \right)^{3/2} r_0^3} \,,
\end{split}
\end{equation}
where $\sigma_\mathrm{ann}$ is the annihilation cross section, $v_\chi$ is the velocity of the dark matter, $n_\chi$ is the DM number density and $r_0$ is defined from the DM distribution, such that:
\begin{equation}
\begin{split}
n_\chi(r) \equiv n_0 \, e^{- r^2/r_0^2} = n_0 \exp \left[ - \frac{m_\chi \Phi(r)}{T_\mathrm{DM}} \right] \,,
\end{split}
\end{equation}
where $n_0 = \frac{N_\chi}{\pi^{3/2} r_0^2}$ is the DM number density at the center of the star, $\Phi(r) \equiv V(r) / m_\chi \simeq \frac{2}{3} \pi G \left(\rho_c + 3 p_c \right) r^2$ is the gravitational potential per unit mass at $r$~\cite{Goldman:1989nd}, $\rho_c$ and $p_c$ are the density and the pressure of the core of the star, $G$ is Newton's gravitational constant, $T_\mathrm{DM}$ is the DM's temperature, and the second equality follows from considering the virial theorem, such that $\langle K \rangle = \langle V \rangle$. Then, by equating those expression for the DM distribution we obtain, 
\begin{equation}
\begin{split}
r_0 \sim \sqrt{\frac{3 T_\mathrm{DM}}{2 \pi G m_\chi \left(\rho_c + 3 p_c \right)}} \,.
\end{split}
\label{eq:riso}
\end{equation}
We note that the radius of the DM thermosphere can be found from $r_{\textrm{th}} = \sqrt{\langle r^2 \rangle} = (3/2)\sqrt{T_{\mathrm{DM}}/\pi G m_\chi \left(\rho_c + 3 p_c \right)} = \sqrt{3/2}\,r_{0}$. 

We finally need to compute the number of DM particles enclosed in the sphere of radius $r_0$. This can be done by considering the full evolution of the number of DM fermions:
\begin{equation}
\begin{split}
\frac{dN_\chi}{dt} = C - A_\chi N_\chi^2\,,
\end{split}
\end{equation}
where $C$ is the capture rate. Therefore, the total number of particles as a function of time is the solution of the previous equation:
\begin{equation}
\begin{split}
N_\chi (t) = \sqrt{\frac{C}{A_\chi}} \tanh \left( \sqrt{C A_\chi} t \right)\,.
\end{split}
\end{equation}

Therefore, we can now compute the contribution to $\dot{E}_k$ in \Cref{eq:E_k_dot} coming from annihilations, which gives us a total energy rate deposition of:
\begin{equation}
\begin{split}
\dot{E}_k^\mathrm{tot} = \dot{E}_k + m_\chi A_\chi N_\chi^2\,.
\end{split}
\end{equation}

\begin{figure}
    \centering
    \includegraphics[width=0.75\textwidth]{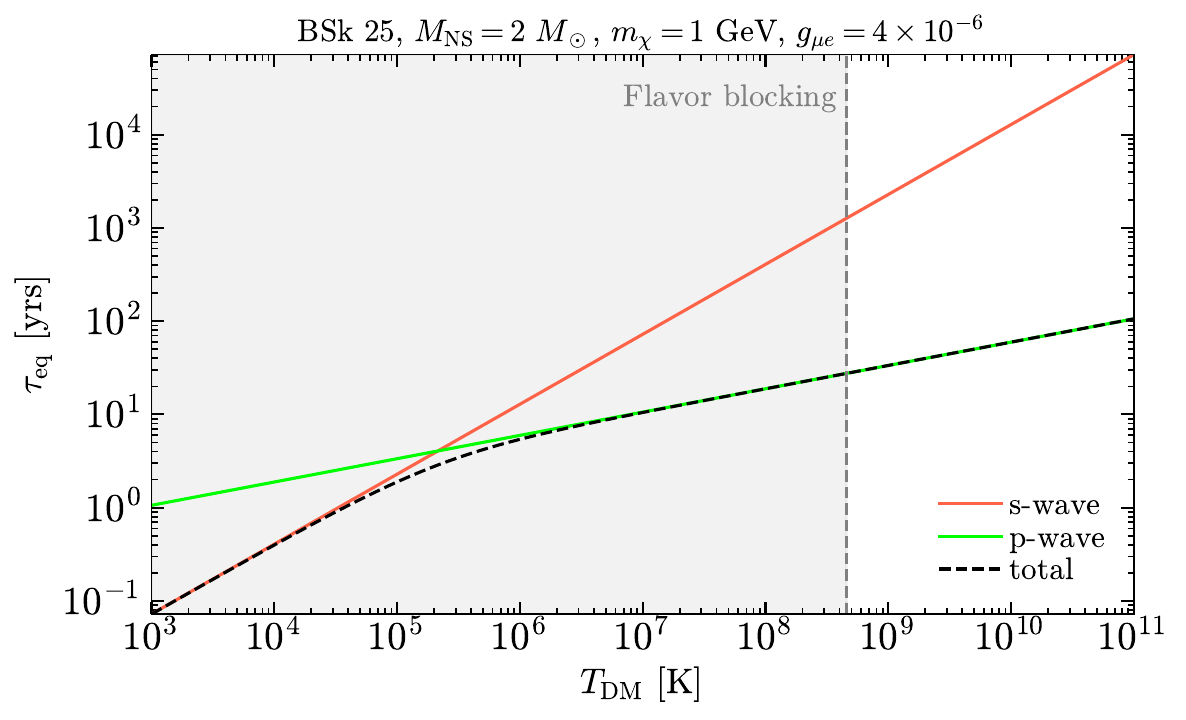}
    \caption{ Equilibration time with respect to each channel and the DM temperature for an NS of mass of $2 \ M_\odot$ (BSk 25) and DM of mass $1$ GeV.
     \label{fig:t_eq}}
\end{figure}

Compared to the age of an old and cold NS, an equilibrium between capture and annihilation is easily attained, such that $C = A_\chi N_\chi^2$, so that $\dot{E}_k^\mathrm{tot} = \dot{E}_k + m_\chi C$. This is driven by the $p$-wave annihilation process, which dominates thanks to the higher temperatures of the DM due to flavor blocking, as can be seen in \Cref{fig:t_eq}.

\bibliographystyle{JHEP}
\bibliography{lib}

\end{document}